\documentclass[%
 superscriptaddress,
 amsmath,amssymb,
 aps,
 pra,
twocolumn,
longbibliography
]{revtex4-2}

\usepackage[whole]{bxcjkjatype} 

\usepackage[unicode]{hyperref}

\usepackage{graphicx}
\usepackage{lineno}
\usepackage{dcolumn}
\usepackage{bm}
\usepackage{braket}
\usepackage{hyperref}
\hypersetup{
  colorlinks=true,
  allcolors = blue
}

\usepackage{xcolor} 
\definecolor{darkblue}{rgb}{0.,0.24,0.51}
\definecolor{britishracinggreen}{rgb}{0.0, 0.26, 0.15}
\definecolor{darkgreen}{rgb}{0,0.60,.2}

\newcommand{\beq}{\begin{equation}}
\newcommand{\eeq}{\end{equation}}
\newcommand{\bk}{{\bf k} }
\newcommand{\bp}{{\bf p} }
\newcommand{\bq}{{\bf q} }

\newcommand{\down}{\downarrow}
\newcommand{\up}{\uparrow}

\begin{document}

\preprint{APS/123-QED}
\title{Stability and breakdown of Fermi polarons in a strongly interacting Fermi-Bose mixture}


\author{Isabella Fritsche}
\affiliation{Institut für Quantenoptik und Quanteninformation (IQOQI), Österreichische Akademie der Wissenschaften, 6020 Innsbruck, Austria}
\affiliation{Institut für Experimentalphysik, Universität Innsbruck, 6020 Innsbruck, Austria}

\author{Cosetta Baroni}
\affiliation{Institut für Quantenoptik und Quanteninformation (IQOQI), Österreichische Akademie der Wissenschaften, 6020 Innsbruck, Austria}
\affiliation{Institut für Experimentalphysik, Universität Innsbruck, 6020 Innsbruck, Austria}

\author{Erich Dobler}
\affiliation{Institut für Quantenoptik und Quanteninformation (IQOQI), Österreichische Akademie der Wissenschaften, 6020 Innsbruck, Austria}
\affiliation{Institut für Experimentalphysik, Universität Innsbruck, 6020 Innsbruck, Austria}

\author{Emil Kirilov}
\affiliation{Institut für Quantenoptik und Quanteninformation (IQOQI), Österreichische Akademie der Wissenschaften, 6020 Innsbruck, Austria}
\affiliation{Institut für Experimentalphysik, Universität Innsbruck, 6020 Innsbruck, Austria}
\author{Bo Huang (黄博)} 
\affiliation{Institut für Quantenoptik und Quanteninformation (IQOQI), Österreichische Akademie der Wissenschaften, 6020 Innsbruck, Austria}
\affiliation{Institut für Experimentalphysik, Universität Innsbruck, 6020 Innsbruck, Austria}

\author{Rudolf Grimm}
\affiliation{Institut für Quantenoptik und Quanteninformation (IQOQI), Österreichische Akademie der Wissenschaften, 6020 Innsbruck, Austria}
\affiliation{Institut für Experimentalphysik, Universität Innsbruck, 6020 Innsbruck, Austria}

\author{Georg M. Bruun}
\affiliation{Center for Complex Quantum Systems, Department of Physics and Astronomy, Aarhus University, Ny Munkegade 120, DK-8000 Aarhus C, Denmark}
\affiliation{Shenzhen Institute for Quantum Science and Engineering and Department of Physics, Southern University of Science and Technology, Shenzhen 518055, China}

\author{Pietro Massignan}
\affiliation{Departament de F\'isica, Universitat Polit\`ecnica de Catalunya, Campus Nord B4-B5, 08034 Barcelona, Spain}

\date{\today}
\begin{abstract}
We investigate the properties of a strongly interacting imbalanced mixture of bosonic $^{41}$K impurities immersed in a Fermi sea of ultracold $^6$Li atoms. This enables us to explore the Fermi polaron scenario for large impurity concentrations including the case where they form a Bose-Einstein condensate. The system is characterized by means of radio-frequency injection spectroscopy and interspecies interactions are widely tunable by means of a well-characterized Feshbach resonance. We find that the energy of the Fermi polarons formed in the thermal fraction of the impurity cloud remains rather insensitive to the impurity concentration, even as we approach equal densities for both species. The apparent insensitivity to high concentration is consistent with a theoretical prediction, based on Landau's quasiparticle theory, of a weak effective interaction between the polarons. The condensed fraction of the bosonic $^{41}$K gas is much denser than its thermal component, which leads to a break-down of the Fermi polaron description. Instead, we observe a new branch in the radio-frequency spectrum with a small energy shift, which is consistent with the presence of Bose polarons formed by $^{6}$Li fermions inside the $^{41}$K condensate. A closer investigation of the behavior of the condensate by means of Rabi oscillation measurements support this observation, indicating that we have realized Fermi and Bose polarons, two fundamentally different quasiparticles, in one cloud.
\end{abstract}

\maketitle

\section{INTRODUCTION}
Quantum many-body systems may greatly vary in the nature of their elementary participants and in energy scales, descending from nuclear and quark-gluon plasmas, electrons in condensed matter, down to liquid helium and ultracold gases. Nonetheless, the theoretical approaches used to tackle them are remarkably similar \cite{Landau1933, Strinati2018tbb, woelfle2018qic, Bowley1973mto, Massignan2014pdm}. One of the most important tools developed to deal with the many-body problem, and to simplify it drastically, is Landau's celebrated idea of quasiparticles \cite{Landau1933}. It turns out that the low energy excitations of a large class of many-body systems can be described in terms of particle-like entities denoted quasiparticles. This leads to a relatively simple yet powerful description of interacting many-body systems, and as a consequence the quasiparticle framework is an indispensable tool in our understanding of nature~\cite{Baym1991lfl}. Indeed, while exotic new materials such as unconventional superconductors \cite{Norman2011} or singular Fermi liquids \cite{Varma2002} may defy this quasiparticle description, Landau's framework has in general been spectacularly successful in describing a wide range of systems in nature. 

Multi-component ultracold gases offer an excellent test bed to investigate quantum many-body systems \cite{Bloch2008mbp}. In particular, strongly imbalanced quantum mixtures represent an ideal system to study the limits of Landau's quasiparticle paradigm. In these systems, the minority component represents impurities interacting with the surrounding majority component to form quasiparticles. Since early experiments in 2009 \cite{Schirotzek2009oof, Nascimbene2009coo}, the case of dilute impurities in a large Fermi sea realizing quasiparticles coined Fermi polarons has been intensively studied in many experiments \cite{Kohstall2012mac, Cetina2015doi, Cetina2016umb, Scazza2017rfp, DarkwahOppong2019ooc, Ness2020ooa, Adlong2020}. Thanks to the flexibility provided by ultracold atom experiments, also the complementary case of Bose polarons, i.e.\ quasiparticles formed by embedding mobile impurities in a bosonic environment, has been investigated \cite{Hu2016bpi, Jorgensen2016ooa,Ardila2019,Yan2020bpn,Skou2020}. 

In the single impurity limit, the quantum statistics of the minority species, i.e.\ whether it is a fermion or a boson, is irrelevant for the behavior of the ensemble. Theoretical predictions based on Landau's approach have shown excellent agreement with experimental observations in this regime \cite{Chevy2010ucp,Massignan2014pdm, Levinsen2015Feb, Schmidt2018umb}. Even for moderate impurity concentration, a description in terms of quasiparticles has proved accurate. However, as the concentration is further increased, the quantum statistics of the impurities will determine the fate of the polaron. In Fermi-Fermi systems the impurities first form a Fermi sea of polarons \cite{Scazza2017rfp}, and finally the whole system undergoes a transition to a paired superfluid as the concentration is increased beyond a critical value for attractive interactions~\cite{Pitaevskii2016book,Kinnunen2018}. In contrast, bosonic impurities at large concentration and low temperature will form a  Bose-Einstein condensate (BEC), as we have shown in previous work \cite{Lous2017toa, Lous2018pti, Huang2019}. Employing resonantly tunable interactions, a strongly interacting Fermi-Bose mixture, embedded in the Fermi sea, can then be created. Furthermore, an intrinsic property of quasiparticles such as polarons is that they interact via density modulations in the surrounding medium~\cite{Baym1991lfl}. Such induced interactions between bosonic impurities will in general be attractive, in contrast to fermionic impurities~\cite{Mora2010,Yu2010,Yu2012,CamachoLandau}, and may lead to the formation of bound dimer states~\cite{Camacho2018bia}.

In this Article, we present our experimental observations regarding polaron physics in Fermi-Bose mixtures, where the bosons ($^{41}$K atoms) represent the minority species immersed in a sea of ultracold fermions ($^6$Li atoms). We explore different density regimes and show that both the Fermi and the Bose polaron can be realized in our system. In Sec.~\ref{sec:ch2} we discuss the basic properties of the impurities as a function of their concentration, and the differences with respect to the previously investigated Fermi-Fermi case of $^{40}$K impurities in a $^6$Li gas~\cite{Kohstall2012mac, Cetina2015doi,Cetina2016umb}. After this we introduce our experimental procedures and the relevant parameters in Sec.~\ref{sec:exp_pro}. Then our experimental results are presented and discussed in Sec.~\ref{sec:ExpRes} before we conclude in Sec.~\ref{sec:fin}.

\section{Bosonic Impurities in a fermionic environment}
\label{sec:ch2}
In this Section, we discuss our basic approach of immersing bosonic potassium atoms, $^{41}$K, as a minority component into a Fermi sea of ultracold lithium atoms, $^6$Li, in the presence of strong interspecies interactions. We introduce the three different density regimes accessible in our system. Then we compare the current experimental approach with our previous work, in which we investigated a system where the impurity was represented by the fermionic isotope $^{40}$K \cite{Kohstall2012mac,Cetina2015doi,Cetina2016umb}.
\subsection{From a single impurity to a BEC}
\begin{figure}
\includegraphics[width=\linewidth]{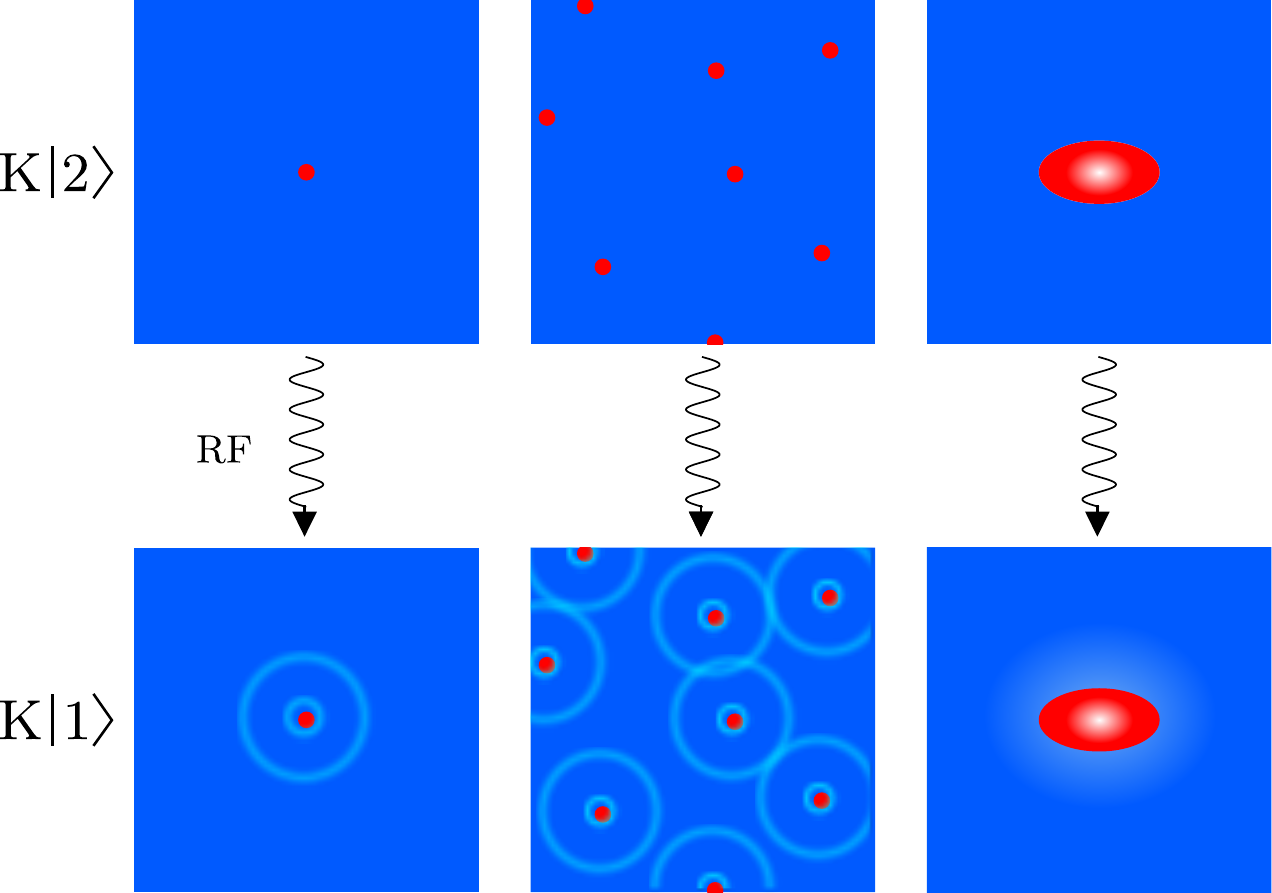}
\caption{\label{fig:exp_setup} Illustration of the Fermi-Bose mixture in three different impurity density regimes. The upper (lower) row shows the non-interacting (interacting) impurities, immersed in a Fermi sea, which is represented by the blue background. The interaction between the impurities and the Fermi sea gives rise to density modulations as illustrated by the light blue circular rings around the K atoms.
A radio-frequency (RF) pulse brings the system from a non-interacting to a strongly interacting state. The three columns illustrate three different regimes. From left to right we increase the bosonic density from a single impurity, to high densities,
and finally to a mixed phase containing a large BEC component.}
\end{figure}
Our main motivation is to investigate density-dependent effects of Fermi polarons emerging from bosonic impurities. The three different regimes of impurity densities in our Fermi-Bose (FB) mixture are illustrated in Fig.~\ref{fig:exp_setup}. The blue background and the red dots represent the Li Fermi sea and the K impurities, respectively. As in our previous work on the Fermi-Fermi (FF) system \cite{Kohstall2012mac}, we use radio-frequency (RF) injection spectroscopy to transfer atoms from a non-interacting spin state K$\ket{2}$ into a state K$\ket{1}$ that interacts with the fermionic medium. 

In the case of a single impurity (left column), the K atom is dressed by particle-hole excitations of the Fermi sea, which lead to local density modulations in the medium and to the formation of the Fermi polaron.  In this low concentration regime, the quantum statistics of the impurity does not matter. The situation is accurately described in terms of a variational ansatz \cite{Chevy2006upd}, which has been widely applied in the field \cite{Massignan2012pad, Massignan2014pdm, Parish2016qdo, Schmidt2018umb}.

As we add more K atoms, we expect to introduce polaron-polaron interactions into our system, as depicted in the middle column of Fig.~\ref{fig:exp_setup}. In this density regime, the spatial overlap of the density modulations around the impurities will result in an effective interaction between the quasiparticles mediated by the fermions, which is attractive due to the bosonic nature of the $^{41}$K atoms \cite{Mora2010,Yu2010,Yu2012,CamachoLandau,Santamore2008fmi,Hu2018afp,Camacho2018bia,Tajima2018mfp, DeSalvo2019oof, Edri2020oos}, see also Appendix~\ref{app:Th}. This effective interaction plays a key role in Landau's quasiparticle theory, but experimental observations in quantum-degenerate gases are still scarce~\cite{Cetina2016umb}.

In the high-density regime (right column), the impurities form a BEC in the center of the trap. As we shall see, the density of this BEC exceeds that of the fermionic density by a large factor of $\sim36$. In this case, the two species interchange their roles and, locally, the Li atoms can be considered as impurities in the K-BEC. Such a scenario is commonly described in terms of Bose polarons \cite{Hu2016bpi, Jorgensen2016ooa}. Therefore, as we vary the K density from a thermal cloud to a BEC, we can realize the transition from a system of Fermi polarons to a system of Bose polarons.
\subsection{Comparison with previous experiments}
Here we discuss the basic situation investigated in our present work in comparison with our previous experiments. The main difference is the change in the quantum statistics of the impurity species, i.e.\ bosonic $^{41}$K atoms instead of fermionic $^{40}$K atoms. The Fermi sea of $^{6}$Li stays essentially the same, only with minor changes of the particular experimental parameters. This similarity enables us to focus on the effects of the quantum statistics of the impurity.
\\
\indent 
The tunability of the interspecies interaction strength in our experiment is given by a Feshbach resonance (FR) \cite{Chin2010fri} between the lowest Zeeman sublevels of K and Li. The parameters characterizing the FR are very similar in the FB and FF case, see App.~\ref{app:FR} and Supp.~Mat.~of Ref.~\cite{Cetina2016umb}. A quantitative difference is the Fermi energy, which in the present case is somewhat lower and therefore modifies the influence of the finite effective range on the interspecies interaction. This fact is taken into account in our theoretical approach, which is presented in detail in App.~\ref{app:Th}.
\\
\indent Another difference between the two systems, which is connected with the FR, is the choice of spin states we work with. In the FF system, we tune the interactions between the lowest and the third-to-lowest spin state of Li and K, respectively. Therefore, dipolar relaxation \cite{Naik2011fri} can lead to decay into lower lying Zeeman sublevels, which is relevant, in particular, if molecules are formed \cite{Jag2016lof}. In the FB case, the interacting atoms occupy the lowest spin channel, which suppresses the two-body process of dipolar relaxation.

When considering few-body processes \cite{Naidon2017epa, Greene2017ufb}, we find that the quantum statistics of the impurities plays a crucial role. In contrast to the FF system, inelastic few-body scattering processes are not suppressed by Pauli blocking in the present case. Therefore, three-body processes involving one fermion (Li) and two bosons (K) can lead to strong resonant losses \cite{Haefner2017rot, Johansen2017tuo}. Other few-body processes, like, e.g., atom-dimer resonances \cite{Jag2014ooa}, sensitively depend on the quantum statistics.

As we increase the K density and generate a BEC, which is only possible if the impurities are bosonic, the character of the whole system changes qualitatively. As described in our previous publications \cite{Lous2018pti,Huang2019, Huang2020bec}, already for moderate repulsive interspecies interactions we enter the regime of phase separation. Here the BEC separates from the Fermi gas and behaves as an almost pure BEC. On the other hand, for moderate attractive interactions the BEC is supposed to undergo collapse \cite{Ospelkaus2006idd, Zaccanti2006cot}.
\\
\indent Owing to the fact that the Li-K mixture offers very similar interaction tunability for $^{40}$K and $^{41}$K, it provides an excellent test bed for investigating the differences between strongly interacting FF and FB systems.
\section{Experimental procedures}
\label{sec:exp_pro}
In this Section, we outline the experimental procedures for preparing a mixture of ultracold $^6$Li and $^{41}$K atoms in the vicinity of an interspecies FR. After describing the preparation of our sample (\ref{sec:preparation}), we introduce experimental parameters relevant for the data analysis (\ref{sec:analquant}) and our method of tuning the interspecies interaction (\ref{sec:interaction}). Finally we explain the RF excitation scheme (\ref{sec:polexc}).

\subsection{Sample preparation and detection}
\label{sec:preparation}
We use an all-optical approach \cite{Spiegelhalder2010aop} to prepare our system in a crossed-beam optical dipole trap (CODT), operated with $1064$-nm light. Following the evaporation and spin preparation scheme described in detail in the Supplemental Material of Refs.~\cite{Lous2018pti,Lous2018PHD}, we obtain a mixture of lithium atoms in the lowest hyperfine spin state Li$\ket{1}$ ($F = 1/2, m_F = 1/2$) and potassium atoms in the second to lowest hyperfine spin state K$\ket{2}$ ($F = 1, m_F = 0$) in thermal equilibrium.

At the end of each experimental cycle we switch off the optical dipole trap, let the atoms expand for an adjustable time and detect them using state-selective absorption imaging. This allows us to image the atoms in two spin states per species for each experimental cycle. Details on the imaging technique and on how to obtain the atom number are provided in the Supplemental Material of Ref.~\cite{Cetina2015doi}.

We conduct our measurements in two different regimes, in which we either prepare a thermal cloud (THC) or a partially condensed cloud (PBEC) of K atoms immersed in a degenerate Fermi sea of Li atoms. We keep the same trap setting for both regimes in order to avoid complications arising from different trap depths and different light shifts of the center of the Feshbach resonance. We achieve this by altering the preparation stage for the PBEC with respect to the THC in two ways. First, we increase the initially loaded atom numbers and second we apply an additional evaporation step where we further ramp down the power of our CODT and slowly (within 1s) recompress it to the initial values in the end. With this procedure, we ensure a two-fold increase in the number of K atoms and thus an increase of the critical temperature for condensation by about $30\%$. The condensed fraction is typically of the order of $\beta\approx0.5$.

The finally prepared system consists of roughly $10^5$ Li$\ket{1}$ and $10^4$ K$\ket{2}$ atoms \footnote{Note that the atom numbers in the PBEC and in the THC slightly differ, due to the different preparation methods.} with temperatures of $T\approx100\,$nK at a magnetic field of $B\approx335\,$G, where the only relevant effect of the weak interaction is the thermalization of the sample with an interspecies scattering length of about $\sim60\,a_0$ \cite{HanTie}, $a_0$ being the Bohr radius. The atoms are trapped in a CODT with radial trap frequencies $\omega_{\text{rad,K}}=2\pi\times227\,\text{s}^{-1}$ and $\omega_{\text{rad,Li}}=2\pi\times382\,\text{s}^{-1}$, as well as axial frequencies $\omega_{\text{ax,K}}=2\pi\times31\,\text{s}^{-1}$ and 
$\omega_{\text{ax,Li}}=2\pi\times49.5\,\text{s}^{-1}$ for K and Li, respectively. The resulting elongated trap has an aspect ratio of $\sim7$ with the weak axis oriented horizontally. The differential gravitational sag \cite{Lous2017toa} amounts to about $3\,\mu$m and can be neglected since the Fermi sea is much larger. These are the initial conditions for all the measurements presented in this Article.
\subsection{Relevant parameters}
\label{sec:analquant}
The procedure for thermometry in our mixture of $^{41}$K and $^{6}$Li atoms is different for the two experimental regimes. In the case of THC we determine the temperature in a standard way by ballistic expansion of the K atoms after releasing them from the trap. In the case of PBEC we follow the approach described in Ref.~\cite{Lous2017toa}, where we release the atoms from the trap to determine the condensate fraction of the K atoms. From this and the known atom numbers and trap frequencies, we calculate the temperature. For a PBEC, this thermometry method proved to be more accurate than the standard ballistic expansion method~\cite{Lous2017toa}. The density profiles of both the degenerate Li Fermi gas and the bosonic K cloud are calculated using standard textbook relations \cite{Pitaevskii2016book}. We neglect small finite-size or interaction corrections for the condensate~\cite{Lous2018pti}.
\\
\indent In order to determine the relevant parameters of our system we take into account that the Fermi pressure acts on the Li atoms, and that the optical potential is about two times deeper for K. This leads to the potassium sample being much smaller than the spatial extent of the lithium cloud, which allows us to treat the latter as an essentially homogeneous environment \cite{Kohstall2012mac}. Since we obtain our spectroscopic signal from the K component, we introduce the K-averaged atom number densities, $\bar{n}_\text{Li}$ and $\bar{n}_\text{K}$, for both species,
\begin{align}
  \bar{n}_\text{Li,K} = \frac{1}{N_\text{K}}\int n_\text{Li,K}(\textbf{r}) n_\text{K}(\textbf{r}) d^3\textbf{r},
\end{align}
with $n_\text{Li,K}(\textbf{r})$ being the local number density at position $\textbf{r}$ of Li and K, respectively. Similarly we define the effective Fermi energy as 
\begin{align}
  \epsilon_\text{F} = \frac{1}{N_\text{K}}\int E_\text{F}(\textbf{r}) n_\text{K}(\textbf{r})d^3\textbf{r},
\end{align}
where the local Fermi energy at position $\textbf{r}$ is given by 
\begin{align}
  E_\text{F}(\textbf{r}) = \frac{\hbar^2(6\pi^2n_\text{Li}(\textbf{r}))^{2/3}}{2m_\text{Li}}.
\end{align}
Finally, we define the effective Fermi wave number as ${\kappa_\text{F} = \sqrt{2m_\text{Li}\epsilon_\text{F}}/\hbar}$. 

In Table~\ref{tab:table1} we present an overview of typical values for important experimental parameters, which we adjust to measure the polaron spectra, as discussed in Sec.~\ref{sec:polspec}. Since such a measurement consists of many individual spectra, the given uncertainties reflect the standard deviation for all spectra. 
We introduce the dimensionless range parameter $\kappa_\text{F}R^*$ \cite{Petrov2004tbp}, which quantifies the character of the Feshbach resonance (open- or closed-channel dominated), and the reduced temperature of the sample ${k_BT/\epsilon_\text{F}}$, where $k_B$ is the Boltzmann constant. The total atom numbers of Li ($N_\text{Li}$) and K ($N_\text{K}$) are listed, and we give the concentrations ${\mathcal{C}_\text{K2} = \bar{n}_\text{K2}/\bar{n}_\text{Li}}$ and ${\mathcal{C}_\text{K2,BEC} = \bar{n}_\text{K2,BEC}/\bar{n}_\text{Li}}$ for the thermal and the condensed part of the non-interacting sample, respectively. Note that in the majority of our measurements, we state the concentration of the non-interacting sample. The value for the interacting case ${\mathcal{C}_\text{K1} = \bar{n}_\text{K1}/\bar{n}_\text{Li}}$ is experimentally not directly accessible because of interaction effects on the spatial distribution and can thus only be estimated in Sec.~\ref{sec:den_var}.
\begin{table}[!htbp]
\caption{\label{tab:table1}Table of experimental parameter values for measurements on the thermal cloud (THC) and partial BEC (PBEC).}
\begin{ruledtabular}
\begin{tabular}{lll}
\textrm{parameter}&
\textrm{THC}&
\textrm{PBEC}
\\
\colrule
$\epsilon_\text{F}$ & $k_B\times930(60)\,$nK & $k_B\times620(50)\,$nK\\
$1/\kappa_\text{F}$ & $4000(130)\,a_0$ & $4800(200)\,a_0$\\
$\kappa_\text{F}R^*$ & $0.57(2)$ & $0.47(2)$\\
$T$ & $130(13)\,$nK & $118(21)\,$nK\\
$k_BT/\epsilon_\text{F}$ & $0.14(1)$ & $0.19(3)$\\
$N_\text{Li}$ & $2.8(2)\times10^5$ & $1.2(1)\times10^5$\\
$N_\text{K}$ & $1.2(1)\times10^4$ & $2.7(3)\times10^4$\\
$\bar{n}_\text{Li}$ & $1.9(2)\times10^{12}\text{cm}^{-3}$ & $1.0(1)\times10^{12}\text{cm}^{-3}$\\
$\bar{n}_\text{K}$ & $0.92(7)\times10^{12}\text{cm}^{-3}$ & $1.4(1)\times10^{12}\text{cm}^{-3}$\\
$\bar{n}_\text{K,BEC}$ & -- & $3.8(1)\times10^{13}\text{cm}^{-3}$\\
$\mathcal{C}_\text{K2}$& $0.61(7)$ & $1.5(5)$\\
$\mathcal{C}_\text{K2,BEC}$~\cite{NoteIF22}& -- & $36(6)$\\
$\beta$& -- & $0.46(7)$
\end{tabular}
\end{ruledtabular}
\end{table}
\subsection{Interaction tuning}
\label{sec:interaction}
An interspecies Feshbach resonance (FR) centered at $B_0 = 335.080(1)\,$G between the atoms in states Li$\ket{1}$ and K$\ket{1}$ ($F = 1, m_F = 1$) enables us to tune the $s$-wave interaction by varying the magnetic field. In Appendix~\ref{app:FR} we report on the accurate determination of $B_0$, including our trap-specific light shift~\cite{Lous2018pti}. This allows us to adjust the interspecies scattering length $a$ according to the relation \cite{Chin2010fri}
\begin{align}
  a = a_\text{bg}\left(1-\frac{\Delta}{B-B_0}\right),
\end{align}
where $\Delta = 0.9487\,$G is the width and $a_\text{bg} = 60.865\,a_0$ is the background scattering length of the Feshbach resonance, as explained in detail in the Supplemental Material of Ref.~\cite{Lous2018pti}.

In order to quantify the interspecies interaction strength in our system we introduce the dimensionless interaction parameter $X = -1/(\kappa_\text{F}a)$. Most of the measurements presented in this Article are conducted in the strongly interacting regime ${(-1\lesssim X\lesssim 1)}$, which raises the question of accuracy and precision in our knowledge of the magnetic field strength. Therefore, we experimentally determined the residual fluctuations around the target value, resulting in a statistical uncertainty of $\sigma_B = 0.5\,$mG, which translates to a corresponding uncertainty $\sigma_X < 0.035$ of the interaction parameter. Furthermore, we observe a slow drift of the magnetic field strength, which we take into account by taking the average value of the magnetic field determined before and after each measurement. We disregard all measurements that exceed a magnetic field drift of $3\,$mG.

The uncertainty in the B field and the fact that our FR is extending over a rather small magnetic field region set the resolution we can achieve for $X$. For this reason, we discretize the variation of the interaction parameter and divide a region between $-1.5 < X < 1.5$ into 12 bins, each having a width of $\sim0.25$. Individual bins in the full spectrum, presented in Fig.~\ref{fig:spec_th}, contain averages of $1$-$4$ measurements.
\subsection{Radio-frequency excitation scheme}
\label{sec:polexc}
In order to probe the spectral function of our K atoms across the Feshbach resonance we use radio-frequency (RF) spectroscopy. There are two main schemes, referred to as ``injection'' and ``ejection'' spectroscopy, which shed light on different aspects of the system \cite{Massignan2014pdm, Liu2020}. We choose the former, in which we transfer the minority atoms from a state that is to a good approximation non-interacting into an interacting state. One advantage of this method is that the system can be transferred to a strongly interacting state that is not necessarily the ground state of the system. It therefore enables us to study the repulsive polaron as a metastable state \cite{Massignan2011,Kohstall2012mac} along with its non-equilibrium evolution.

The system is excited by an RF pulse that transfers atoms from the non-interacting K$\ket{2}$ to the interacting K$\ket{1}$ state in the presence of Li$\ket{1}$. In order to avoid side lobes in the spectrum we use a Blackman-shaped pulse. We adjust it to be a resonant $\pi$-pulse for a bare K cloud, i.e.\ in the absence of the Li atoms. The power is chosen such that at the resonance frequency $\nu_0$, where the maximum transfer occurs, we have a pulse duration of $\tau_\text{RF} = 1\,$ms. This duration was chosen as a compromise between spectral resolution and lifetime. The former is set by the spectral width of the RF pulse $\sigma_\text{RF}=0.7\,$kHz, which, depending on the specific sample preparation, is around $\sigma_\text{RF}\approx0.04\,\epsilon_\text{F}/\hbar$. The latter is given by the shortest lifetime of the polaron, which we estimated to be around $1\,$ms.

The presence of Li changes the frequency of maximum transfer because of interactions between the two species. In most of our measurement we vary the frequency detuning $\Delta\nu = \nu_0 - \nu$, keeping the pulse power unchanged, and observe the transferred fraction of potassium atoms $N_\text{K1}/N_\text{tot}$, where $N_\text{K1}$ is the atom number in the K$\ket{1}$ state and $N_\text{tot} = N_\text{K1} + N_\text{K2}$ is the total atom number in both states. The dependence of the spectroscopic signal $N_\text{K1}/N_\text{tot}$ on $\Delta\nu$ reflects the energy spectrum of our strongly interacting system of K$\ket{1}$ atoms immersed in a Li$\ket{1}$ Fermi sea. We determine the uncertainty of the atom numbers from the standard deviation of repeated measurements. A small non-zero background, especially in the PBEC regime, may be attributed to imaging artefacts \cite{noteIF20} and is directly subtracted from the data.
\section{Experimental results}
\label{sec:ExpRes}
In this Section, we present our experimental observations. In Sec.~\ref{sec:polspec}, we discuss our RF measurements of the spectral response of the K atoms. Following this, we describe in Sec.~\ref{sec:den_var} our findings on the energy of the repulsive Fermi polaron as we vary the density of the thermal K atoms of a partial BEC. The emergence of Bose polarons in the condensed component is discussed in Sec.~\ref{sec:BosePol}. Then we present our observations on the lifetime of the repulsive Fermi polaron and discuss possible decay channels in Sec.~\ref{sec:liftime}, before we finally examine the behavior of the K atoms in the PBEC regime on the basis of Rabi oscillation measurements in Sec.~\ref{sec:bec_part}.
\subsection{Spectral response}
\label{sec:polspec}
\begin{figure*} 
  \includegraphics[width = 0.49\linewidth]{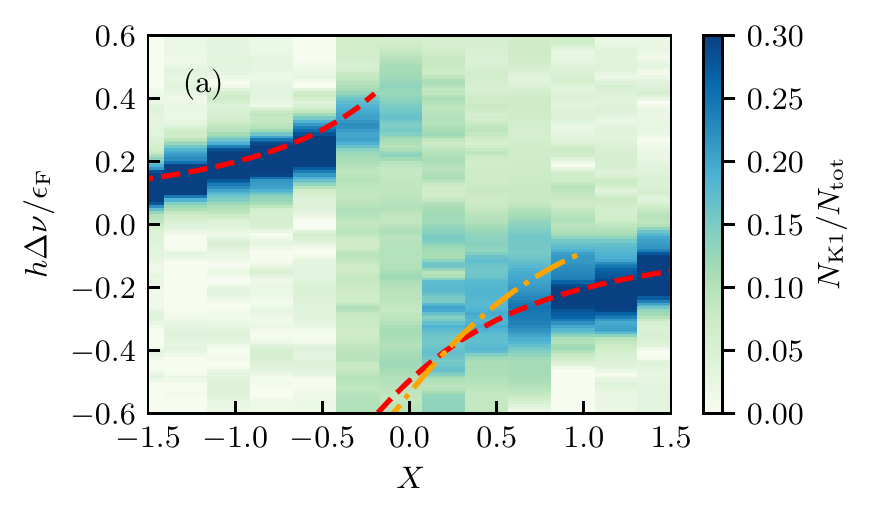}
  \includegraphics[width = 0.49\linewidth]{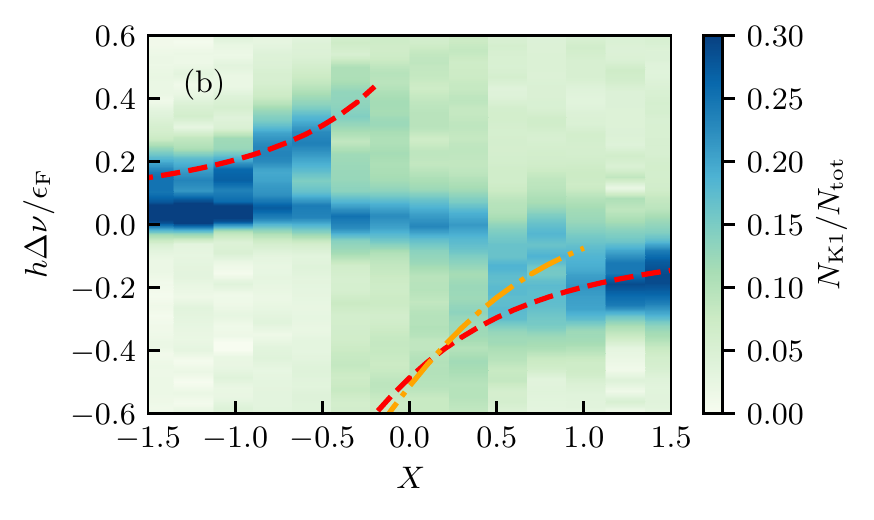}
  \caption{\label{fig:spec_th} Spectral response of a bosonic $^{41}$K sample immersed in a $^6$Li Fermi sea. Panels (a) and (b) show the measured excitation spectra in the thermal cloud (THC) regime and the partially condensed (PBEC) regime, respectively. The spectra are shown as a function of the interaction parameter $X=-1/(\kappa_\text{F}a)$ and the dimensionless RF detuning $h\Delta\nu/\epsilon_F$. The color map refers to the transferred fraction of atoms from K$\ket{2}$ to K$\ket{1}$. Red dashed and orange dash-dotted lines illustrate our theoretical predictions for the polaron and molecule energies in the single-impurity limit, respectively.}
\end{figure*}
\begin{figure*}
  \includegraphics[width = 0.49\linewidth]{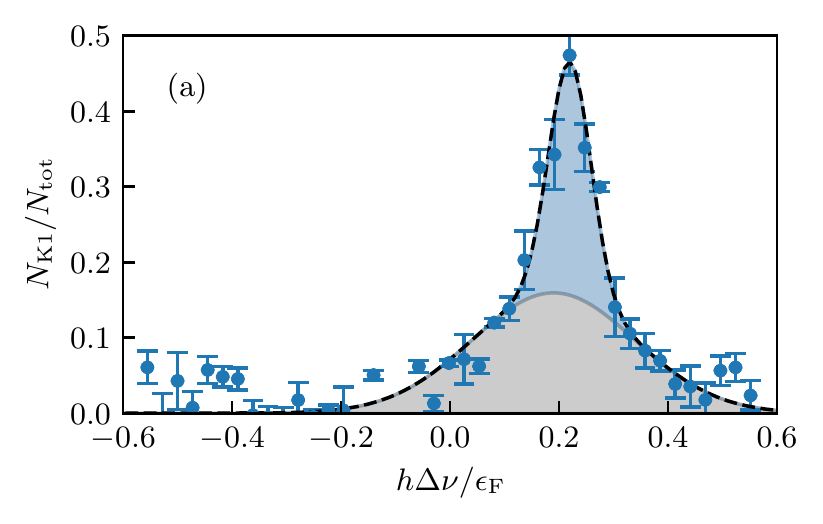}
  \includegraphics[width = 0.49\linewidth]{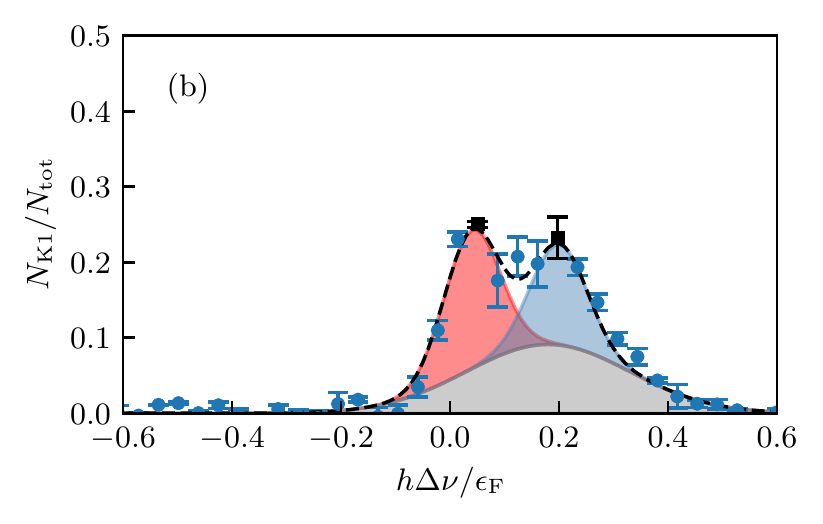}
  \caption{\label{fig:one_spec} Typical polaron spectra in the THC (a) and PBEC (b) regime. We show the fraction of transferred atoms as a function of the frequency detuning $\Delta\nu$ of the applied RF pulse at an interaction strength of $X \approx -0.7$. The shaded areas under the curves illustrate the contributions resulting from a fit by a double-Gaussian (THC, left) and triple-Gaussian (PBEC, right) function. Black dashed lines depict their sum. The width of the narrow peaks is fixed to the Fourier width of the applied pulse. The measurement points marked by black squares in (b) are further investigated in Fig.~\ref{fig:Fig3Inset}.}
\end{figure*}
\begin{figure}
  \includegraphics[width=\linewidth]{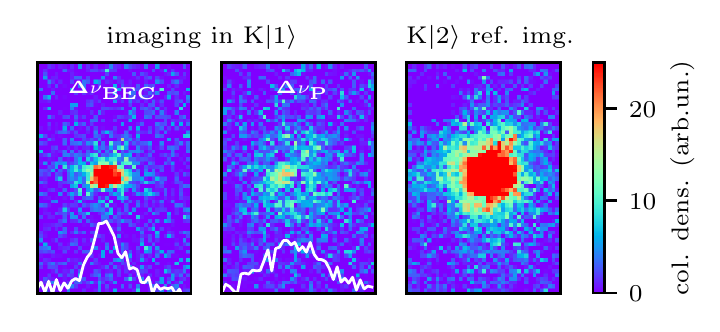}
  \caption{\label{fig:Fig3Inset} Absorption images of K$\ket{1}$ corresponding to the two measurements, marked by black squares in Fig.~\ref{fig:one_spec}(b) and a K$\ket{2}$ reference image after a short time of flight of $6\,$ms and $4\,$ms, respectively. The left panel and the middle panel show the atoms at the transfer frequencies $\Delta\nu_\text{BEC}$ and $\Delta\nu_\text{p}$, respectively. The color map depicts the column density in arbitrary units. The field-of-view of all images is about $(150\times230)\,\mu$m. The white solid lines show the corresponding projected line-density profiles. A reference picture of the K$\ket{2}$ cloud, before transfer, with $\beta\approx0.5$ is shown in the panel on the right.
  }
\end{figure}
In order to investigate the full spectral response of the system across the strongly interacting regime, we combine RF spectra taken at different values of $X$. The spectra, recorded in the thermal (THC) and partially condensed (PBEC) regime, are depicted in Figs.~\ref{fig:spec_th}(a) and (b). The $x$-axis represents the discretized dimensionless interaction parameter $X$, as discussed in Sec.~\ref{sec:interaction}. Each bin shows the transferred fraction $N_\text{K1}/N_\text{tot}$ as a function of the energy detuning of the RF pulse $h\Delta\nu$ normalized to $\epsilon_\text{F}$. 
The theoretical predictions, red dashed and orange dash-dotted lines, denote a variational calculation describing a single impurity interacting with a Fermi sea using a two-channel model \cite{Kohstall2012mac,Massignan2012pad}. The dimensionless range parameter in the two regimes is $\kappa_\text{F}R^* = 0.57(2)$ and $\kappa_\text{F}R^* = 0.47(2)$, respectively.

In Fig.~\ref{fig:spec_th}(a) we show the full spectral response in the THC regime. We observe a typical polaron spectrum consisting of the repulsive and the attractive branch exhibiting a positive and negative energy shift, respectively, and a decrease of contrast as the interaction is tuned close to $X = 0$. The obtained polaron energies are in good agreement with the theoretical predictions for the single impurity scenario, represented by the red dashed lines, although the concentration $\mathcal{C}_\text{K2} = 0.61(7)$ in this measurement is fairly high.

Figure~\ref{fig:spec_th}(b) shows the spectrum in the PBEC regime. A striking difference between the THC and PBEC spectra is that in the latter a new branch, which shows almost no energy shift, emerges in the spectrum. The bimodal spectral response is a consequence of different resonance frequencies of the transfer to the K$\ket{1}$ state for the two components of the gas. The thermal part of the K cloud appears to behave like in the single impurity limit, even though the K density is similar to the Li density. In stark contrast to this, the condensed part is transferred at a frequency close to the non-interacting value $\Delta\nu = 0$, with a small but consistent upshift corresponding to a few percent of the Fermi energy. As we discuss in Sec.~\ref{sec:BosePol}, this shift can be attributed to the formation of Bose polarons, where the Li atoms are now the impurities. 

In order to further investigate the differences between the THC and PBEC regimes we show two sample spectra at an interaction strength of ${X\approx-0.7}$ in Figs.~\ref{fig:one_spec}(a) and (b), respectively. In the THC regime we find a single narrow peak, which we attribute to the Fermi polaron, along with a broader pedestal, which we interpret as a many-body continuum of states. The observed spectrum can be well approximated by a double Gaussian fit $G_{\rm p}(\Delta\nu) + G_{\rm bg}(\Delta\nu)$, as also used in our previous work \cite{Kohstall2012mac}. 
The function takes the form ${G_{\alpha}(\Delta\nu) = A_{\alpha}e^{-(\Delta\nu-\Delta\nu_\alpha)^2/(2\sigma{_\alpha}^2)}}$, with $A_{\alpha},\,\Delta\nu_\alpha,\,\sigma{_\alpha}$ representing the amplitude, center and width of the Gaussian for $\alpha =$ p, bg. The polaron peak, ${\alpha = \text{p}}$, is fixed to a spectral pulse width of $\sigma_p = 0.7\,\text{kHz} \approx 0.04\,\epsilon_\text{F}/h$, which corresponds to the Fourier width resulting from the finite duration of the $1$-ms RF pulse. The background, $\alpha =$ bg, is marked by the gray, broad Gaussian. We transfer about $50\%$ of the atoms into the interacting state at a frequency detuning corresponding to $h\Delta\nu\approx0.2\epsilon_F$.

In the PBEC regime, depicted in Fig.~\ref{fig:one_spec}(b), we identify a maximum transfer at two well-defined frequencies. We approximate the lineshape of the whole spectrum by a triple-Gaussian function. The first two parts stem from the polaron and the many-body continuum ${\tilde{G}_\text{p}(\Delta\nu) + \tilde{G}_\text{bg}(\Delta\nu)}$. We assume that the ratio of the two amplitudes stays the same as determined in Fig.~\ref{fig:one_spec}(a), but their absolute values are reduced corresponding to the fraction of non-condensed atoms, as ${\tilde{G}_\text{p,bg}(\Delta\nu) = G_\text{p,bg}(\Delta\nu)\times(1-\beta)}$. The third part describes the transfer of the condensed fraction ${\tilde{G}_\text{BEC}(\Delta\nu) = G_\text{BEC}(\Delta\nu)\times\beta}$ at a small 
energy shift.

In the two panels on the left of Fig.~\ref{fig:Fig3Inset}, we show absorption images of atoms in K$\ket{1}$, after a short time of flight of $6\,$ms, which were released from the trap within $\sim10\,\mu$s after the RF pulse. The two pictures correspond to the measurements for the two frequency detunings $\Delta\nu_\text{BEC}$ and $\Delta\nu_\text{p}$, for which we have observed maximum transfer of the BEC and the thermal cloud, respectively. These two detunings are marked by black squares in Fig~\ref{fig:one_spec}(b). The atomic clouds in the images have the same atom number, but very different spatial distributions. The left panel shows a dense cloud that only extends over about $40\,\mu$m, whereas the middle panel shows dilute atoms that are distributed over the whole picture. In the right image we present a reference picture of a K$\ket{2}$ cloud before transfer, with a BEC fraction of $\beta\approx0.5$. Comparing these images shows that a fraction of the non-condensed part is transferred in the middle picture and a fraction of the condensed part is transferred in the left picture. This strongly supports our interpretation that the two different frequencies correspond to the resonance frequencies of the two components of the partial BEC.

To conclude this part, our observations show that the spectra for the THC and the non-condensed part of the PBEC sample are consistent with a theoretical description of the Fermi polaron, and with our previous measurements on the Fermi polaron with fermionic impurities~\cite{Kohstall2012mac}. In contrast, the condensed part of the partial BEC, which has a very large concentration of K atoms with $\mathcal{C}_\text{K2}\approx36$, shows a much smaller energy shift that seems unrelated to the Fermi polaron.

\subsection{Concentration variation}
\label{sec:den_var}
\begin{figure}
  \includegraphics[width=\linewidth]{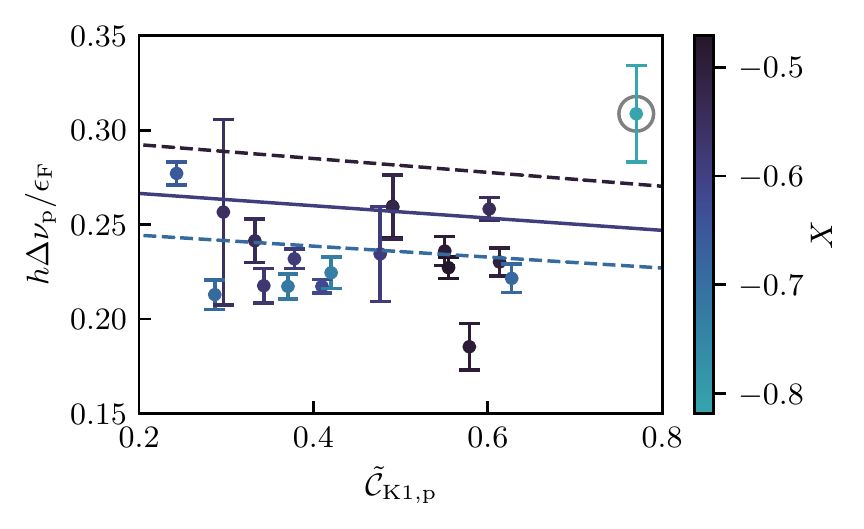}
  \caption{\label{fig:den_var1} Concentration dependence of the energy of the repulsive Fermi polaron. The color code refers to the interaction strength $X$. The solid and dashed lines show the theoretical predictions including polaron-polaron interactions, according to Eq.~\eqref{energyShiftPerImpurity}, for the mean value and its standard deviation of $X = -0.6\pm0.1$. The grey circle marks the measurement with the highest temperature $k_\text{B}T/\epsilon_\text{F}=0.27$ (see discussion in text).}
\end{figure}
We now investigate closer the effects of a finite impurity concentration. In particular, we expect on general grounds that there are interactions between the polarons, which should show up as a change in their energy as a function of their concentration~\cite{Baym1991lfl}. In order to explore this, we take a set of spectra for densities in the range $0 < \mathcal{C}_\text{K2} < 45$ at an interaction strength $X=-0.6(1)$. Here, the uncertainty denotes the standard deviation that characterizes typical experimental fluctuations. We vary the concentration by changing various parameters such as the loading time and the evaporation endpoint in our preparation sequence.

Since only a fraction of the atoms in K$\ket{2}$ is transferred and only atoms in K$\ket{1}$ can be responsible for interaction effects, the concentration $\mathcal{C}_\text{K1}$ is the relevant parameter. This, however, cannot be obtained directly because of our incomplete knowledge of interaction effects on the spatial distribution during the RF pulse. We therefore introduce estimated concentrations, obtained by multiplying the concentration of K$\ket{2}$ by the estimated transferred fraction at the resonance frequency. The measurements presented here are conducted in the PBEC regime and we can therefore obtain two concentrations ${\mathcal{\tilde{C}}_{\text{K1,p}} = \mathcal{C}_\text{K2}\times(\tilde{A}_\text{p}+\tilde{A}_\text{bg})/(1-\beta)}$ and ${\mathcal{\tilde{C}}_{\text{K1,BEC}} = \mathcal{C}_\text{K2}\times\tilde{A}_\text{BEC}/\beta}$ for the non-condensed and the condensed component of the K-atoms, respectively. The amplitudes $\tilde{A}_\alpha$ correspond to the fitting amplitudes, as discussed in Sec.~\ref{sec:polspec}.

In Fig.~\ref{fig:den_var1} we show our results regarding the density variation of the energy of the repulsive polaron. The color scale indicates the particular values of the interaction parameter $X$ for
each data point. From Fermi liquid theory we know that there is an effective interaction $f$ between the polarons mediated 
by the Fermi gas~\cite{Baym1991lfl}. 
As shown in Ref.~\cite{Yu2012} (see also App.~\ref{app:Th}), the effective interaction has a direct and an exchange contribution.
For low temperature and arbitrary Bose-Fermi interaction strength, it can be calculated from the density of states $\mathcal N$ at the Fermi surface of the Li atoms and from the number $\Delta N$ of Li atoms  in the dressing cloud of the polaron as 
${f=-\Delta N^2/\mathcal{N}+g_1}$. Here, $g_1=4\pi\hbar^2 a_{11}/m_K$ represents the direct interaction between two K$|1\rangle$ atoms, where $a_{11}$ is the corresponding scattering length.
Note that the induced interaction ${-\Delta N^2/\mathcal{N}}$, mediated by the Fermi gas, is attractive since the K atoms are bosonic. Taking into account that the RF injection spectroscopy gradually increases the impurity concentration, so that the signal is averaged from zero to the final K density, the observed average energy shift is $E(n)=E(0)+fn/2$, see App.~\ref{app:Th} for details. The lines in Fig.~\ref{fig:den_var1} are obtained from this formula where the solid and dashed lines correspond to an interaction strength of $X=-0.6$ and $\pm$ its standard experimental deviation of $0.1$. We note that due to the small value of the scattering length $a_{11}$ between the atoms in K$|1\rangle$, the negative slope of these lines is essentially only due to the mediated interaction $-\Delta N^2/\mathcal{N}$.

From Fig.~\ref{fig:den_var1} we see that our experimental observations are consistent with the predicted concentration dependence of the polaron energy. The mean temperature of the measurements presented is ${k_\text{B}T/\epsilon_\text{F} = 0.17(2)}$ so that we expect the result to be fairly close to the zero-temperature limit assumed by the theory. The measurement marked by the grey circle has an exceptionally high temperature of $k_\text{B}T/\epsilon_\text{F} = 0.27$. We therefore suspect that this data point is subject to a significant finite-temperature shift and may thus be considered an outlier.

Given the large fluctuations in the data and the predicted small influence of the effective interaction, we cannot provide conclusive evidence of its presence. Instead, the comparison shows that future improved experiments may indeed open up the possibility to observe the effect of polaron-polaron interactions, for which a clear observation is still missing in the field of ultracold quantum gases.

\subsection{Bose polarons}
\label{sec:BosePol}
\begin{figure}
  \includegraphics[width=\linewidth]{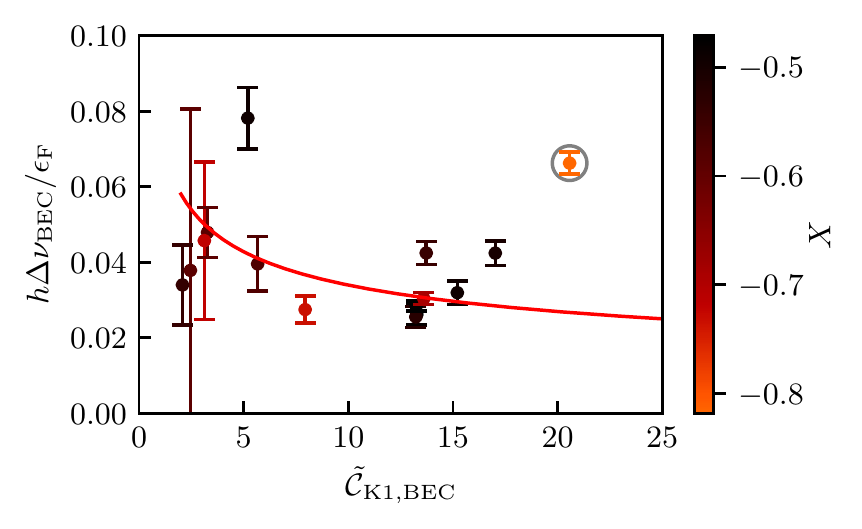}
  \caption{\label{fig:den_var} Concentration dependence of the observed BEC peak position. The color code refers to the interaction strength $X$ and the grey circle marks the measurement with highest temperature, as in Fig.~\ref{fig:den_var1}. A fit to the data of the theoretical prediction according to Eq.~\eqref{eq:bosepolaron} is shown as the red solid line. The error bars of the data points represent the uncertainties of the fits.}
\end{figure}
We now turn to the low-energy peak, which, as we have shown, comes from the condensed fraction of the K atoms. In Fig.~\ref{fig:den_var}, the position of this peak is shown as a function of the impurity concentration, extracted from the same dataset as presented in Sec.~\ref{sec:den_var}. We observe a small and consistent energy shift of $\sim0.04$. An estimation of this energy shift may be obtained as follows.

First, since the three scattering lengths between the K atoms in the two spin states ($a_{11}$, $a_{22}$ and $a_{12}$) differ by less than $0.3\%$ \cite{TiemannPrievComm}, the energy shift must be attributed mostly to K-Li interactions. Second, since the density of the condensed part of the K atoms is much higher than for the Li atoms in the center of the trap, the situation is reversed in the sense that one can now regard the Li atoms as impurities in a BEC of K atoms. A suitable framework to analyze this is therefore the  one of Bose polarons, formed by Li atoms in the K$|1\rangle$ BEC, rather than the one of Fermi polarons. The total energy shift 
can therefore be estimated as $\Delta E_{\rm tot} = N_{\rm Li} E_{\rm Li}$, where $N_{\rm Li}$ is the number of Li atoms inside the K$|1\rangle$ BEC, and $E_{\rm Li}$ is the energy of a single Bose polaron. In the strongly interacting region on the BEC side of the resonance ($X\approx -0.6$), a repulsive Bose polaron has a typical energy $E_{\rm Li}=\xi\epsilon_n$, where $\xi$ is a constant of order unity~\cite{Hu2016bpi, Jorgensen2016ooa,Ardila2019,Yan2020bpn,Skou2020}. The energy scale $\epsilon_n$ of the Bose gas is defined, in analogy with the Fermi energy, as $\epsilon_n=\hbar^2 \kappa_n^2 / (2 m_{\rm K})$, with $\kappa_n=(6 \pi \bar{n}_\text{K1,BEC})^{1/3}$ so that $\epsilon_n/\epsilon_{\rm F} = (m_{\rm Li}/m_{\rm K})(\mathcal{C}_\text{K1,BEC})^{2/3}$. The relevant concentration is that of the K$|1\rangle$ BEC that interacts with the Li atoms, which we approximate as $\mathcal{C}_\text{K1,BEC}\approx\mathcal{\tilde{C}}_\text{K1,BEC}$. Since RF spectroscopy measures the energy shift per atom transferred from K$|2\rangle$ to K$|1\rangle$, the relevant quantity is the energy shift 
per K atom in the  K$|1\rangle$ BEC, which is given by 
\begin{equation}
\label{eq:bosepolaron}
  \Delta E_{\rm tot} / N_{\rm K1,BEC} = (6/41) (\mathcal{\tilde{C}}_\text{K1,BEC})^{-1/3} \xi \epsilon_{\rm F}.
\end{equation}
With $\xi$ as the only free parameter, Eq.~\eqref{eq:bosepolaron} can then be fitted to the experimental data displayed in Fig.~\ref{fig:den_var}, which yields ${\xi\approx0.5}$. The  resulting curve, shown by the solid line in Fig.~\ref{fig:den_var}, reasonably agrees with the data. We should however mention a few caveats. First, the K$|1\rangle$ BEC is only formed above a certain critical concentration, but the RF probe transfers the atoms gradually into the K$|1\rangle$ state. This effect is further explored in Sec.~\ref{sec:bec_part}. It follows that the observed behavior is presumably a result of an average BEC density experienced by the Li atoms during the RF probe. Second, the bosons and the fermions will eventually phase separate for the given interaction strength \cite{Lous2018pti,Huang2019}, which also complicates the interpretation of the experiment. Nevertheless, the agreement between theory and experiment for a reasonable value of the fit parameter, ${\xi\approx0.5}$, suggests that the observed shift of the BEC energy is, indeed, due to the formation of Bose polarons in the center of trap. %
\subsection{Lifetime of repulsive polaron}
\label{sec:liftime}
\begin{figure}
\includegraphics[width=\linewidth]{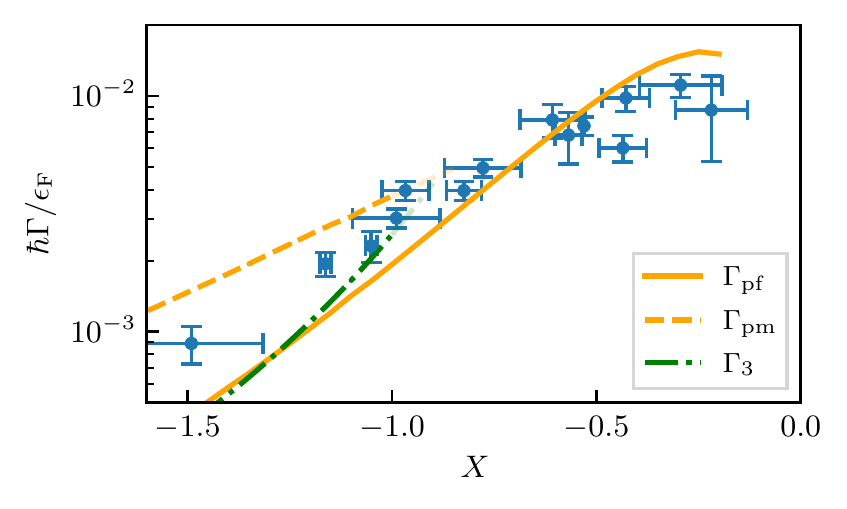}
\caption{\label{fig:lifetime_pol} Decay rate of the polaron for different interaction strengths $X$. Blue circles depict the measured lifetimes of the polaron. The orange solid and dashed lines show theoretical calculations of the two- and three-body decay, respectively. The three-body recombination rate in vacuum is depicted by the green dash-dotted line. See App.~\ref{app:Th} for details.}
\end{figure}
\begin{figure}
\includegraphics[width=\linewidth]{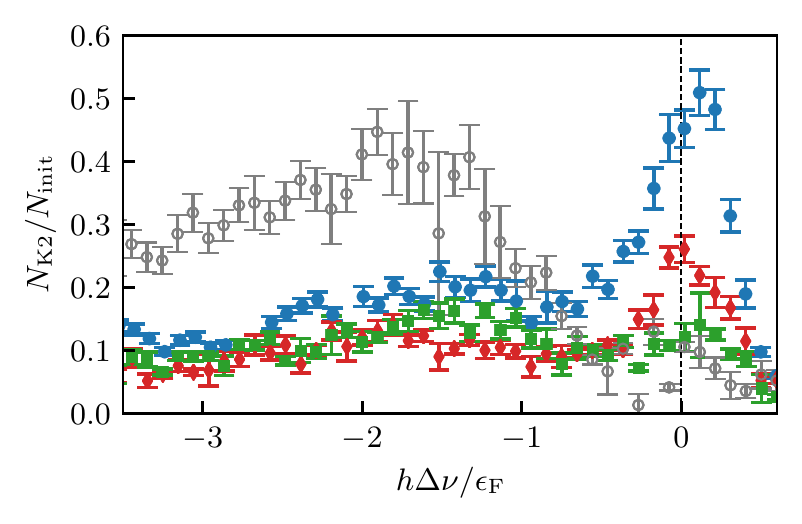}
\caption{\label{fig:molproof} Ejection spectra of the repulsive polaron and its decay products. We show the fraction of atoms transferred from K$\ket{1}$ to K$\ket{2}$ as a function of the applied RF signal. The blue circles, red diamonds, green squares show the spectrum after a decay time of $1.2\,$ms, $2.2\,$ms, $5.2\,$ms. These three measurements are normalized to the total atom number of the measurements with the shortest wait time (blue circles). As a comparison, we also show a molecule spectrum (gray empty circles).}
\end{figure}
The repulsive Fermi polaron is a metastable quasiparticle, which can decay via two- or three-body processes into lower energy states~\cite{Massignan2011,Massignan2014pdm}. In order to determine its lifetime, we carry out measurements in the THC regime for $X < 0$. The repulsive polaron is populated by applying a $\pi$-pulse with a duration $\tau = 0.3\,$ms (instead of the $1\,$ms used in all measurements shown before) and frequency detuning $\Delta\nu_\text{p}$. In this way, we resonantly excite the quasiparticle with a short pulse in order to maximize the number of transferred atoms. After this excitation, about ${\sim50\%}$ of the atoms are found to remain in K$\ket{2}$. We therefore apply a $10\,\mu$s resonant ``cleaning'' light pulse to remove them from the trap, thus creating a pure sample of strongly interacting K$\ket{1}$ and Li$\ket{1}$ atoms. At this point, we wait for a variable time before applying another RF pulse, identical to the first one, which only addresses the polarons that have not yet decayed. In contrast to all measurements presented so far, the measured signal is now the fraction of atoms transferred back into the non-interacting state K$\ket{2}$. We fit an exponential decay to the data sets obtained for various values of $X$ and extract the $1/e$ decay time $\tau_\text{p}$, which represents the lifetime of the polaron. 

The blue circles in Fig.~\ref{fig:lifetime_pol} show the repulsive polaron decay rate ${\Gamma = 1/\tau_\text{p}}$ as a function of the interaction strength. Approaching the resonance, the decay rate rises from ${10^{-3}\,\epsilon_\text{F}/\hbar}$ at $X = -1.5$ to about ${10^{-2}\,\epsilon_\text{F}/\hbar}$ at $X\approx-0.2$. This corresponds to polaron lifetimes between $\sim10\,$ms and $\sim1\,$ms and is in excellent agreement with our previous experiments on Fermi polarons with fermionic impurities \cite{Kohstall2012mac}. 

The solid line in Fig.~\ref{fig:lifetime_pol} is a theoretical prediction based on the assumption that the repulsive polaron decays via a two-body process into the attractive polaron, which due to its high kinetic energy can be approximated by a free particle. The dashed line gives, on the other hand, the three-body decay rate into the molecule, taking into account medium effects in the perturbative regime. Finally, the green dash-dotted line shows the three-body decay rate in a vacuum for a broad resonance~\cite{Petrov2003tbp}, adapted here to describe a narrow resonance. For details on the calculations of these rates, see App.~\ref{app:Th}. By comparing these theory lines with the data, we see that two-body decay into the attractive polaron seems to be the main loss channel for strong interactions. However, for weaker interactions the attractive polaron is ill-defined, due to the smallness of its residue and decay into the molecular states. In this regime, three-body decay processes become dominant. This is consistent with the observations for the case of fermionic impurities~\cite{Kohstall2012mac,Scazza2017rfp}.

We observe a residual signal remaining in K$\ket{1}$ after the second RF pulse, which transfers the repulsive polarons into K$\ket{2}$. It consists of remaining polarons and its decay products. In order to investigate the nature of the residual component, we let the polaron decay for a time $t$ and then we apply ejection spectroscopy. In contrast to the measurement described so far, we now vary the frequency of the second RF pulse, which transfers K$\ket{1}$ atoms back to K$\ket{2}$. In Fig.~\ref{fig:molproof} we show such measurements for the three decay times $t_1 = 1.2\,$ms, $t_2 = 2.2\,$ms, and $t_3 = 5.2\,$ms, all taken at the same interaction strength $X=-0.80(2)$. We show the transferred fraction $N_{K2}/N_\text{init}$ normalized to the total atom number ${N_\text{init} = N_\text{K1}(t_{1}) + N_\text{K2}(t_{1})}$ after a wait time of $t_1$. The blue circles, red diamonds, and green squares represent the ejection spectra recorded after waiting times of $t_1$, $t_2$, and $t_3$, respectively.

We expect that the decay product consists of  molecules, since this is the predicted ground state for $X=-0.80$. In order to check this, we compare the ejection spectra with a molecule dissociation spectrum, shown by the gray open circles in Fig.~\ref{fig:molproof}. To obtain this spectrum, we start with a THC sample in the non-interacting state. Then we associate molecules by applying a $3\pi$ pulse to K$\ket{2}$ at a frequency adjusted such that it corresponds to the binding energy of the molecule at $X = -0.80$ (see App.~\ref{app:FR}). Since we do not transfer all K$\ket{2}$ atoms into the molecular state, we apply a resonant ``cleaning'' light pulse, which removes the remaining atoms from the trap. This leaves us with a mixed sample of Li$\ket{1}$-K$\ket{1}$ molecules and bare Li$\ket{1}$ atoms. Then we perform ejection spectroscopy to probe the spectrum of the molecule. This is achieved by applying another $3\pi$ pulse to dissociate the molecules, where we vary the frequency. Note that this particular spectrum is normalized to its own total atom number $N_\text{init} = N_\text{tot}$.

Let us now compare the four ejection spectra presented in Fig.~\ref{fig:molproof}. In the measurement at the shortest decay time (blue circles) we recognize a narrow peak at positive energies, which we identify as the repulsive polaron. The broad pedestal at negative energies on the other hand reflects the response of the molecules, since it is similar to the bare molecular spectrum. As we increase the wait time from $t_1$ to $t_2$ and then to $t_3$, we observe a decrease of transferred atoms at the repulsive polaron frequency, as a consequence of its decay. 

Given that the polarons decay into molecules, we would expect a corresponding increase in their spectral signal, i.e.\ the broad pedestal. This is however not observed. Instead, as the wait time of the measurements in Fig.~\ref{fig:molproof} is increased from $t_1$ to $t_3$, we see a reduction of K atoms in the trap by a factor of ${\sim2}$, while the broad pedestal is unchanged. From this, and the measurements presented in Fig.~\ref{fig:lifetime_pol}, we speculate that the repulsive polarons decay into molecules, which themselves undergo relatively fast collisional decay into lower lying molecular states, where the excess energy of the latter is sufficient to remove the atoms from the trap. We believe Bose-Fermi dimers are less robust against collisions as compared to Fermi-Fermi dimers, for which we have demonstrated a Pauli suppression effect in Ref.~\cite{Jag2016lof}.
\subsection{Rabi oscillation measurements}
\label{sec:bec_part}
\begin{figure}
\includegraphics[width=\linewidth]{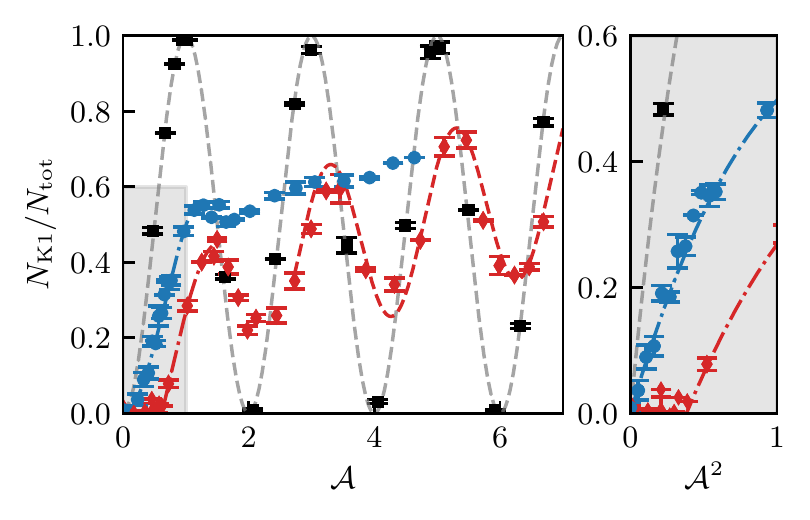}
\caption{\label{fig:Rabi_tot} Rabi oscillation measurements. We show the dependence of the transferred fraction on the pulse area, normalized to a $1\,$-ms RF $\pi$-pulse in the non-interacting case. The black squares show the Rabi oscillations of a non-interacting sample. The two further measurements are conducted in the THC and the PBEC regime at $\nu_p$ (blue circles) and $\nu_\text{BEC}$ (red diamonds), respectively. The black and red dashed lines show a $\sin^2$ oscillation at the non-interacting Rabi frequency, where the latter has a reduced amplitude by the factor $\beta$, corresponding to the BEC fraction. The blue and red dash-dotted curve show the initial transient of a $\sin^2$ oscillation with Rabi frequencies reduced by the interaction. Right-hand panel: the region of weaker RF pulses, marked by the shaded area, is plotted against the square of the pulse area $\mathcal{A}^2$.}
\end{figure}
We now further investigate the nature of the thermal and condensed parts of the K cloud by performing Rabi oscillation measurements, as shown in Fig.~\ref{fig:Rabi_tot}. A $1$-ms RF pulse is applied to transfer atoms from K$\ket{2}$ into K$\ket{1}$. The transferred fraction of atoms is then measured as a function of the pulse area $\mathcal{A} = \sqrt{P/P_\pi}$, where the peak RF power $P$ of our Blackman pulse is the experimentally controlled variable and $P_\pi$ is the corresponding power to achieve a $\pi$-pulse in a non-interacting case. First, we take a reference measurement with Li removed from the trap. As we vary the RF power the black squares show the Rabi oscillations of the non-interacting sample, which are well fitted with a $\sin^2$ function, as illustrated by the black dashed line. After this, we prepare our atoms in the THC at $X\approx-0.5$ and tune the radio frequency to the polaron peak at $\nu_\text{p}$ (blue circles). We observe an initial increase in the signal that follows a $\sin^2$ behavior (dash-dotted blue line). For $\mathcal{A}\gtrsim1$, this changes into a steady increase in the transferred fraction with no clear oscillations. We can explain this effect by the decay of the polaron to other states, such as molecules \cite{Massignan2012pad}. Such states have a reduced overlap with the non-interacting state. Therefore, the transfer probability from K$\ket{2}$ to K$\ket{1}$ is higher than the backtransfer from the dressed molecular state to K$\ket{2}$. This results in a growing population in K$\ket{1}$ with increasing RF power.

When we prepare a PBEC sample and tune the frequency of the RF pulse to $\nu_\text{BEC}$ (red diamonds) the system behaves in a very different way. In the region $\mathcal{A}\gtrsim1$ of Fig.~\ref{fig:Rabi_tot}, we observe a clear oscillating behavior, depicted by the red dashed line. The frequency is the same as for the non-interacting case, but the amplitude is reduced by a factor that is close to the BEC fraction $\beta$. This is consistent with a BEC of K atoms oscillating between the $|1\rangle$ and $|2\rangle$ states, in a way, largely unaffected by the small concentration of Li atoms. The increasing background can be attributed to an off-resonant contribution originated from the non-condensed component.

A remarkable feature shows up in the behavior of the condensate for weak RF pulses. For $0<\mathcal{A}\lesssim1$, we find that the atom transfer is inhibited. In order to highlight this striking effect, we plot the transferred fraction in the region of small pulse areas, marked by the shaded region, as a function of $\mathcal{A}^2$ in the right panel of Fig.~\ref{fig:Rabi_tot}. This representation turns an initial quadratic dependence on $\mathcal{A}$, typical for the coherent evolution of a quantum system, into a linear depence on $\mathcal{A}^2$. Such a behavior is nicely visible in all three data sets. However the red diamonds show a transfer of the BEC only after a critical value of $\mathcal{A}^2\approx0.4$ is reached.

This peculiar effect likely arises from a density-dependent shift of the resonance frequency. In the regime of low concentration $\mathcal{C}_\text{K1}$ the final state of the system is the Fermi polaron. This results in almost no transfer, for small $\mathcal{A}$ in Fig.~\ref{fig:Rabi_tot}, since the detuning of the RF pulse to the polaron energy is about $\sim4\Gamma_p$, where $\Gamma_p$ is the spectral width of the polaron peak and the Fourier width of the RF pulse is $1/\tau_\text{RF}\approx\Gamma_p$. On the other hand, when the RF pulse transfers enough atoms to create a K$|1\rangle$ BEC, the resonance frequency shifts to the one determined by the Bose polarons and permits the transfer to start.

On top of this effect, as the BEC density increases in K$\ket{1}$, phase separation may occur and can remove the fermions from the spatial region occupied by the bosons \cite{Lous2018pti,Huang2019}. We estimate this effect to take place while the RF pulse is applied, since there is no clear separation of the corresponding time scales in our experiment. In this scenario, the two species will separate at an RF power that is high enough for a significant 
fraction of the BEC to be transferred. After this, the K cloud will exhibit Rabi oscillations similar to the non-interacting case.

The origin of the observed inhibition of Rabi oscillations of an RF-coupled BEC in the environment of a Fermi sea is an interesting many-body phenomenon and needs further investigation in future work.
\section{Summary and conclusion}
\label{sec:fin}
We have presented first observations concerning the Fermi polaron with bosonic impurities and its differences with respect to fermionic impurities. The quantum-statistical nature of the impurities, which does not matter in the single-particle limit, enters the problem at higher concentration and can profoundly change the properties of the system. We have explored the case of high densities below and above the threshold for Bose-Einstein condensation of the impurity cloud and found very different behavior.

For a thermal impurity cloud we have probed the energy of the attractive and the repulsive quasiparticle branch across the strongly interacting regime and found properties very similar to those of the previously investigated Fermi-Fermi system. Our observations are, within the experimental uncertainties, fully consistent with the single-impurity theoretical predictions despite the fact that the concentration is near unity.

In order to increase the impurity concentration we have cooled the sample further to create a partial BEC. The spectral response of this dense system reveals a drastic change of the spectrum. We find that, in addition to the signature of the repulsive and attractive polaron, a new branch, the BEC branch, emerges in the spectrum, which shows no sign of the Fermi polaron anymore. Instead we find a small positive shift in energy over a wide range of interactions.
We speculate that, since the concentration far exceeds unity, this effect may be explained by an interchange of the role of the two atomic species, where the BEC and the Fermi sea represent the environment and the impurities, respectively. Such a scenario is usually described by the Bose polaron \cite{Hu2016bpi, Jorgensen2016ooa}. This suggests that the Fermi and the Bose polaron appear as different branches of one spectrum.

We have dedicated particular attention to the region of positive scattering lengths, where the repulsive Fermi polaron is realized. As we vary the concentration, the energy shift of the condensed component of the partial BEC remains small and positive. We find good qualitative agreement with a Bose polaron description, where the back action of the Bose polarons on the surrounding results in a small, but clearly observable energy shift.

As we investigate the concentration dependence of the thermal component of the partial BEC closer, at strong repulsive interactions, our results indicate a slightly smaller energy  of the Fermi polaron than expected from a single-impurity prediction. The experimental uncertainty in the determination of the interaction strength, which is very sensitive to magnetic field fluctuations, renders a qualitative analysis impossible. However, theoretical calculations, including polaron-polaron interactions, predict a decreasing energy shift with increasing concentration, which is consistent with our experimental data. This suggests that interaction effects amongst polarons could be observed in future more precise measurements.

In order to further characterize the metastable repulsive Fermi polaron, we have measured its decay rate and compared it to theoretical predictions of different decay channels. Our observations close to the center of the FR are in very good agreement with two-body scattering processes, where the repulsive polaron decays into a bare particle. Furthermore we find qualitative agreement between the measured decay rates for moderate interactions and our theoretical calculations of three-body decay.

In order to gain further insight into the transitional behavior from low to high concentration, we vary the strength of the spectroscopy pulse that transfers the partial BEC into the state strongly interacting with the fermionic medium. For low pulse strengths, we observe a peculiar interaction-induced inhibition of the transfer, whereas for high pulse strengths we essentially recover the behavior of a non-interacting cloud. This striking result suggests a shift of the resonance frequency with changing concentration, which supports our interpretation of a transition of our mixture between regimes governed by two fundamental quasiparticles, the Fermi and the Bose polaron.

Our capability of creating a partial BEC, which interacts strongly with a surrounding Fermi sea, allows us to investigate the behavior of vastly different concentration regimes, in the same setup. Future measurements focused on the transition between the two fundamentally different polarons could shed light on the largely unknown physics beyond the single quasiparticle picture, where polaron-polaron interactions play a significant role. Conducting measurements on that order of precision will require even better magnetic field control and more stable conditions, which seems feasible with further technical improvements. In addition, time-domain methods \cite{Cetina2015doi, Cetina2016umb} may provide deeper insight into density-dependent behavior. The unambiguous observation of such effects would represent a  major step, since effective interactions are an integral part of Landau's theory of quasiparticles leading to many of its non-trivial predictions.
\begin{acknowledgments}
We acknowledge fruitful discussions with Richard Schmidt and his group at the Max-Planck-Institute for Quantum Optics, and with Jook Walraven. We acknowledge financial support within the Doktoratskolleg ALM (W1259-N27), funded by the Austrian Science Fund FWF, and within the Innsbruck Laser Core Facility, financed by the Austrian Federal Ministry of Education, Science and Research. G.M.B.~acknowledges support from DNRF through the Center for Complex Quantum Systems (Grant agreement No.~DNRF156) and the Independent Research Fund Denmark-Natural Sciences (Grant No.~DFF-8021-00233B). P.M.~further acknowledges support by the Spanish MINECO (FIS2017-84114-C2-1-P), and EU FEDER Quantumcat.
\end{acknowledgments}

\newpage
\appendix
\section{Accurate determination of the Feshbach resonance center}
\label{app:FR}
\begin{figure}
\includegraphics[width=\linewidth]{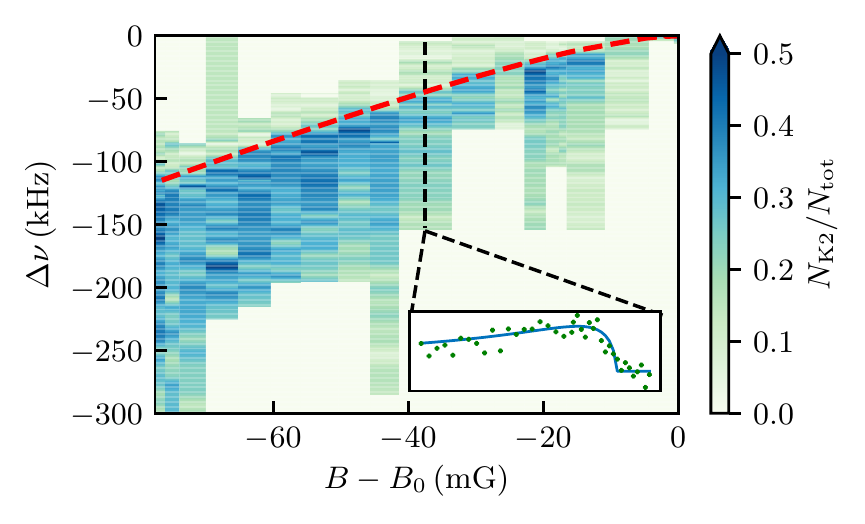}
\caption{\label{fig:B0_vac_disso} Molecule dissociation measurement. The transferred fraction to K$\ket{2}$ (color scale) is shown in dependence of magnetic field and radio-frequency detuning for various magnetic fields around the center of the Feshbach resonance. The inset shows an example spectrum, taken at $B-B_0\approx-37\,$mG. After determining the binding energy of the molecules as a function of the interaction strength we fit the resulting data by Eq.~(\ref{eq:binding}), shown by the red dashed line, with $B_0$ as the only fitting parameter.}
\end{figure}
In all measurements described in this manuscript we tune the inter-particle interaction by means of a particular Feshbach resonance (FR). The exact determination of the Feshbach resonance center $B_0$ of the FR on the mG scale is crucial in our experiment since the strongly interacting regime is only about $\pm12\,$mG wide. Our previous determinations of the FR parameters are explained in detail in the Supplemental Material of \cite{Lous2018pti}. Following a slightly different technique, we determine the molecule's binding energy $E_b$, in vacuum, at different interaction strengths and fit it with 
\begin{align}
\label{eq:binding}
  E_b = \frac{\hbar^2}{8(R^*)^2m_r}\left(\sqrt{1-\frac{4R^*(B-B_0)}{a_{bg}\Delta}}-1\right)^2,
\end{align}
derived in \cite{Petrov2004tbp, Levinsen2011ada}. The reduced mass ${m_r=m_Fm_B/(m_F+m_B)}$, the resonance width $\Delta=0.9487\,$G, the range parameter \cite{Petrov2004tbp} $R^*=2241(7)\,a_0$, and the background scattering length $a_{bg}=60.865\,a_0$ are known, which leaves $B_0$ as the only fitting parameter in this model. It is important to note, that Eq.~\eqref{eq:binding} describes the binding energy of molecules in vacuum. Interactions with the remaining Fermi sea are not included and can lead to a systematic shift on the order of $5\,$mG. 

In contrast to our previous method (see Appendix of \cite{Lous2018pti}), $B_0$ is determined by molecule dissociation \textit{in vacuum} via ejection spectroscopy. The measurement consists of the creation of molecules in the THC regime at $X\approx-0.8$ by applying an RF pulse to the K$\ket{2}$ atoms at a frequency that corresponds to the binding energy $\nu_{RF}\approx E_b/h$, which we optimize roughly on maximum molecule association efficiency. Then, we ramp the magnetic field to lower $X$ values and therefore increase the binding energy. This procedure prevents the molecules from dissociating as we apply an RF pulse to transfer the remaining unbound Li$\ket{1}$ atoms into Li$\ket{2}$. To be sure that no particles, except the molecules, are present we apply a $10\,\mu$s resonant cleaning pulse to Li$\ket{2}$ and another one to K$\ket{2}$. Then we ramp back the magnetic field to reach the final interaction strength $X$ for which we want to determine the binding energy. At this field we apply another RF pulse to transfer the K1 atoms into K2 and consequently dissociate the molecules. As we vary the frequency of this last pulse, we obtain the dissociation spectrum with a line shape determined by the frequency-dependence of the Franck-Condon factor, as described in Ref.~\cite{Chin2005rft}. The inset of Fig.~\ref{fig:B0_vac_disso} shows a sample spectrum at a magnetic detuning $B-B_0\approx-37\,$mG and the corresponding fit to extract the binding energy. We record dissociation spectra in a range of about $80\,$mG where we expect molecules to exist. These measurements are presented in Fig.~\ref{fig:B0_vac_disso}. A fit to the binding energies following Eq.~(\ref{eq:binding}) is illustrated by the red dashed line. The resulting value for the center of the Feshbach resonance is $B_0 = 335.080(1)\,$mG. Note that this value refers to our particular trap setting and includes a light shift of about $25\,$mG. All experiments reported here were carried out for the same trap setting, so that we have the same $B_0$ for all our measurements.


\section{Quasiparticle properties of Fermi polarons}
\label{app:Th}
In this Appendix, we present the calculations from which we infer the quasiparticle properties of isolated Fermi polarons and the strength of their mutual interactions. 

\subsection{Properties of isolated polarons}
A single impurity with momentum $\bp$ immersed in a homogeneous Fermi sea may be described as a quasiparticle, whose dressing is composed of a superposition of particle-hole excitations in the Fermi sea. Such a state can be accurately modeled by the variational Ansatz 
\cite{Chevy2006upd}
\begin{equation}\label{1PH_ansatz}
|\psi\rangle= \phi c^\dagger_{\bf p\downarrow}|\text{FS}\rangle + \!
\sum_{q<k_F}^{k>k_F}\phi_{\bk \bq} c^\dagger_{{\bf p}+{\bf q}-{\bf k}\downarrow}c^{\dag}_{{\bf k}\uparrow}\,c_{{\bf q}\uparrow}\,|\text{FS}\rangle
\end{equation}
Here $c_{\mathbf p\sigma}^\dagger$ creates an impurity ($\sigma=\downarrow$) or a majority atom ($\sigma=\uparrow$) with momemtum $\mathbf p$, and $|\text{FS}\rangle$ denotes the unperturbed Fermi sea. We warn the reader that, to avoid clutter, the notation adopted here slightly differs from the one used in the manuscript.

The  minimization of the energy based on this variational Ansatz yields an identical result to the diagrammatic calculation within the ``ladder" (or ``forward-scattering") approximation \cite{Combescot2007nso}, but the latter (once properly analytically continued \cite{Massignan2011}) allows also to investigate the properties of the repulsive branch, and eventually the effects of non-zero temperature, in a straightforward way.
The retarded self-energy of a single impurity of mass $m_{\downarrow}$, with momentum \bp and energy $\omega$ in a Fermi sea of particles with mass $m_{\uparrow}$ reads ($\hbar=k_B=1$) 
\begin{align}\label{self_energy}
\Sigma(\mathbf{p},\omega)&=\sum_{\bq} f(\xi_{\bq\up}) T(\bp+\bq,\omega+\xi_{\bq\up})\\
&=\sum_{\bq}\frac{f(\xi_{\bq\up})}
{\frac{m_r}{2\pi \tilde{a}}-\sum_{\bk}\left[\frac{1-f(\xi_{\bk\up})}{\omega-(\epsilon_{\mathbf{p+q-k}\down}+\epsilon_{k\up}-\epsilon_{q\up})+i0_+}+\frac{2m_r}{k^{2}}\right]},
\end{align}
where $f(x)=1/[\exp(\beta x)+1]$ is the Fermi function at inverse temperature $\beta$, and $T(\mathbf{P},\Omega)$ is the $T$-matrix describing the scattering of an $\up\down$ pair of atoms with total momentum $\mathbf{P}$ and total energy $\Omega$. Here we introduced the kinetic energy of a $\sigma$ atom measured with respect to the chemical potential $\xi_{\bk\sigma}=\epsilon_{\bk\sigma}-\mu_\sigma=k^2/2m_\sigma-\mu_\sigma$, the reduced mass $m_r=m_{\uparrow}m_{\downarrow}/(m_{\uparrow}+m_{\downarrow})$, and the energy-dependent quantity 
\beq
\frac{1}{\tilde{a}(\omega, K)}=\frac{1}{a}+R^*k_r^2,
\eeq
where $k_r=\sqrt{2m_r[\omega-K^2/(2M)+E_F]}$ (with $K=|\bp+\bq|$ and $M=m_\up+m_\down$) is the relative momentum of the colliding pair, and $E_F$ is the Fermi energy of the homogeneous Fermi sea.
Since we consider the properties of a single $\down$ particle, we have set its chemical potential to zero.

The Green's function of the impurity reads
\beq
G_\down(\bp,\omega)={1 \over \omega-\epsilon_{\bp\down}-\Sigma(\bp,\omega)+i0_+}.
\eeq
Its spectral function $A=-2{\rm Im}[G_\down]$ features two branches of excitations, one at negative and one at positive energies.
In the vicinity of these sharp excitations, the  Green's function at small momenta may be approximated as
\beq
G_\down(\bp,\omega)
\approx\frac{Z_\pm}{\omega-E_\pm-\frac{p^2}{2m^*_\pm}-i Z_\pm{\rm Im}[\Sigma(\bp,E_{\pm})]}.
\eeq
The energy of an attractive $(-)$ polaron at zero momentum is the purely real solution at negative energies of 
\begin{equation}\label{aAttr}
E_{-}=\Sigma(0,E_{-}),
\end{equation}
while the energy of the repulsive $\bp=0$ polaron is the positive energy solution of
\begin{equation}\label{eRep}
E_{+}=\mathrm{Re}[\Sigma(0,E_{+})].
\end{equation}
The quasiparticle residues $Z$ are defined as
\beq
Z_\pm=\frac{1}{1-\mathrm{Re}[\partial_\omega\Sigma(0,\omega=E_{\pm})]},
\eeq
and the effective masses are given by
\beq
m^*_\pm= {m_\downarrow/Z_\pm \over 1+\mathrm{Re}[\partial_{\epsilon_{\down\bp}}\Sigma(0,E_{\pm})]}.
\eeq
The energy, residue, and effective mass  obtained in this way compare very favorably with both MC simulations and experiments \cite{Prokofev2008,Schirotzek2009oof,Nascimbene2009coo,Kohstall2012mac,Koschorreck2012,Scazza2017rfp}.
The energies of dressed molecules are instead computed from a related Ansatz, describing a bare molecule dressed by particle-hole excitations in the medium~\cite{Mora2009gso,Punk2009,Combescot2009,Massignan2012pad,Trefzger2012,Qi2012hpf}.

\subsubsection{Polaron decay}
The repulsive polaron is unstable towards decay into lower-lying excitations, but it remains a well-defined quasi-particle as long as its decay rate $\Gamma$ is small \cite{Bruun2010,Massignan2011,Kohstall2012mac, Scazza2017rfp}.
The population decay rate for the 2-body  process leading a polaron to decay onto free particles (pf) is given by
\beq
\Gamma_{\rm pf}=-2Z_+ {\rm Im}[\tilde\Sigma(0,E_{\downarrow +})],
\eeq
where $\tilde \Sigma$ is defined in Eq.\ (S.16) of Ref.~\cite{Scazza2017rfp}. The competing process leading a polaron to decay onto a dressed molecule is instead given by \cite{Kohstall2012mac}
\begin{align}
\Gamma_{\rm pm}=&\frac{64 k_Fa}{45\pi^3}(Z^3_+Z_M)\left(\frac{m_\up}{m_+^*}\right)^2\left(1+\frac{m_\up}{m_\up+m_\down}\right)^{3/2}\nonumber\\
&\times \left(\frac{E_F}{E_+-E_M}\right)^{5/2}\frac{a}{a^*\sqrt{1+4R^*/a^*}}E_F,
\end{align}
where $E_M$ is the energy of a dressed molecule (found by a variational Ansatz \`a la Chevy), and $a^*=\sqrt{2m_r E_b}$ is the typical size of a vacuum dimer at a narrow resonance.

In the extreme BEC limit, where medium effects become negligible, and in presence of a broad resonance, the three-body recombination proceeds at a rate~\cite{Petrov2003tbp} 
\beq\label{Gamma3Petrov}
\Gamma_3 = \left(\frac{\bar\epsilon_\up}{\epsilon
}
\right)\alpha n_\up^2.
\eeq
Here, $\bar\epsilon_\up$ is the average kinetic energy of majority atoms, $\epsilon$ is the binding energy of the $\up\down$ dimer, and $\alpha$ is a constant which for our mass ratio takes the value
\beq
\alpha_\up = 2.57 \frac{ \hbar^5}{m_\up^3\epsilon^2}.
\eeq
We plot for comparison this formula in Fig.~\ref{fig:lifetime_pol}, using for the majority kinetic energy the $T=0$ value $\bar\epsilon_\up=3E_F/5$, and for the dimer binding energy its value at a narrow resonance, given by Eq.~\eqref{eq:binding}.

\subsection{Polaron-polaron interactions}
An intrinsic property of quasiparticles is that they interact. 
Within Fermi liquid theory \cite{Mora2010,Yu2010,Yu2012}, the total energy density of a gas containing $N_\down\ll N_\up$ impurities in a large sea of $N_\up$ ideal fermions may be written as 
\beq
\mathcal{E}(n_\up,n_\down)=\frac{3}{5}E_F n_\up+E_\down n_\down+\frac{1}{2}fn_\down^2.
\eeq 
The first term in this expression represents the energy of the unperturbed Fermi sea, the second is the contribution of isolated polarons, and the third term is the polaron-polaron interaction. We have neglected the mean kinetic energy of the impurities, which is expected to be very small when impurities are bosonic.

The effective interaction $f$ between Landau quasiparticles contains two contributions: $f=g_1+f_{\rm x}$. The first one is the direct (or mean-field) interaction,  $g_1=4\pi\hbar^2a_{11}/m_{\downarrow}$, where $a_{11}$ is the scattering length between bare impurities. 
The second term instead describes an exchange contribution, mediated by particle-hole excitations in the Fermi sea. At $T=0$, this induced interaction between bosonic impurities is given by~\cite{Yu2012}
\beq\label{fLandau}
f_{\rm x}=-\frac{(\Delta N)^2}{\mathcal{N}}.
\eeq
Here $\mathcal{N}=\frac{3n_\up}{2E_F}$ is the density of states at the Fermi energy, and $\Delta N$ is the number of particles in the dressing cloud of a polaron, given by \cite{Massignan2011} 
\beq
\Delta N
\equiv\left.\frac{\partial n_\uparrow}{\partial n_\downarrow}\right|_{\mu_\uparrow}
=-\left(\frac{\partial \mu_\down}{\partial n_\up}\right)_{n_\down}/\left(\frac{\partial \mu_\up}{\partial n_\up}\right)_{n_\down}
\approx-\frac{\partial \mu_\down}{\partial E_F}.
\eeq
In the last step, we used that $\mu_\up\approx E_F$.

We present here a compact derivation of Eq.~\eqref{fLandau}, following the lines of the elegant presentation given in Ref.~\cite{Yu2012}. Within Landau theory, a $\up$ atom and a $\down$ {\it polaron} interact with a coupling constant $g_{\rm x}$ given by 
\beq
g_{\rm x}=\frac{\partial^2\mathcal{E}}{\partial n_\up\partial n_\down}=\frac{\partial \mu_\up}{\partial n_\down}.
\eeq
To second order in $g_{\rm x}$, the polaron-polaron interaction is then given by 
\beq\label{E2}
\mathcal{E}^{(2)}
=-\frac{g_{\rm x}^2}{V^3}\sum_{\bp_\up,\bp_\down,\bq}
\frac{(1-f_{\bp_\up+\bq})(1+f^{\rm (b)}_{\bp_\down-\bq})f^{\rm (b)}_{\bp_\down}f_{\bp_\up}}{\frac{(\bp_\up+\bq)^2}{2m_\up}+\frac{(\bp_\down-\bq)^2}{2m^*_\down}-\frac{p^2_\down}{2m_\down^*}-\frac{p^2_\up}{2m_\up}},
\eeq
where $f^{\rm (b)}$ indicates Bose functions since we are assuming a bosonic impurity.
The exchange contribution to Landau's polaron-polaron interaction can be calculated from this as 
\beq
f_{\rm x}=\frac{\delta^2\mathcal{E}^{(2)}}{\delta f^{\rm (b)}_{\bp_\down}\delta f^{\rm (b)}_{\bp_\down-\bq}}
\eeq
where both $\bp_\down$ and $\bq$ are vanishingly small.
This gives 
\beq
f_{\rm x}=-\frac{g^2}{V}\left(\sum_{\bp_\up}\frac{f_{\bp_\up}-f_{\bp_\up+\bq}}{\frac{(\bp_\up+\bq)^2}{2m_\up}-\frac{p_\up^2}{2m_\up-}}\right)_{q\rightarrow 0}=g_{\rm x}^2\chi,
\eeq
where $\chi$ is the so-called Lindhard function.
At zero temperature, $\chi$ equals simply the density of states at the Fermi surface $\mathcal{N}=\frac{\partial n_\up}{\partial \mu_\up}=\frac{3n_\up}{2E_F}$. Collecting the above results, at zero temperature we have
\begin{align}
f_{\rm x}&=-g_{\rm x}^2\mathcal{N}
=-\left(\frac{\partial \mu_\up}{\partial n_\down}\right)^2 \frac{\partial n_\up}{\partial \mu_\up} 
= -\left[-  \frac{\left(\frac{\partial \mu_\up}{\partial n_\down}\right)}{\left(\frac{\partial \mu_\up}{\partial n_\up}\right)}\right]^2\frac{\partial \mu_\up}{\partial n_\up}\nonumber\\
&= -\frac{\left(\Delta N\right)^2}{ \mathcal{N}}.
\end{align}
In the last step we used 
$
\left(\frac{\partial x}{\partial y}\right)_z \left(\frac{\partial y}{\partial z}\right)_x \left(\frac{\partial z}{\partial x}\right)_y=-1$.

When the impurities are fermionic, an almost identical calculation leads to $f_{\rm x}^{\rm (f)}=-f_{\rm x}$. Physically, this comes from the Pauli repulsion between identical fermions or alternatively, because the effective interaction involves the exchange of the impurities, which leads to a sign change for fermions as compared to bosons~\cite{Mora2010,Yu2010,Yu2012,CamachoLandau}. 

The Landau interaction $f$ between bosonic impurities is finally given by
\beq
f=-\frac{(\Delta N)^2}{\mathcal{N}}+g_{1}.
\eeq
Note that the Landau polaron-polaron induced interaction (which is the first term in the latter expression) is always attractive for bosonic impurities (and repulsive for fermionic ones), irrespective of whether the impurity-bath interaction is attractive or repulsive.

Introducing the impurity concentration $\mathcal{C}=n_\down/n_\up$, 
the increase of the energy of the gas when adding one impurity is found to be
\beq
\mu_\down = \frac{\partial \mathcal{E}}{\partial n_\down}
= E_\down 
-\frac{2}{3} (\Delta N)^2 \,\mathcal{C}\,E_F 
+g_{1}n_\down.
\eeq 
In RF injection, we are gradually increasing the number of impurities, and therefore the polaron-polaron interactions. Taking a simple average, one gets
\beq
\bar\mu_\down=\frac{1}{N_\down}\int_0^{N_\down}\mu_\down(N_\down')\, dN_\down'=\Delta E,
\eeq
where $\Delta E$ is the energy shift per impurity
\beq\label{energyShiftPerImpurity}
\Delta E = \frac{\mathcal{E}-\frac{3}{5}E_F n_\up}{n_\down}
= E_\down 
-\frac{1}{3} (\Delta N)^2 \,\mathcal{C}\,E_F 
+\frac{g_{1}n_\down}{2}.
\eeq 
\bibliography{apssamp}

\providecommand{\noopsort}[1]{}\providecommand{\singleletter}[1]{#1}%
\begin{thebibliography}{77}%
\makeatletter
\providecommand \@ifxundefined [1]{%
 \@ifx{#1\undefined}
}%
\providecommand \@ifnum [1]{%
 \ifnum #1\expandafter \@firstoftwo
 \else \expandafter \@secondoftwo
 \fi
}%
\providecommand \@ifx [1]{%
 \ifx #1\expandafter \@firstoftwo
 \else \expandafter \@secondoftwo
 \fi
}%
\providecommand \natexlab [1]{#1}%
\providecommand \enquote  [1]{``#1''}%
\providecommand \bibnamefont  [1]{#1}%
\providecommand \bibfnamefont [1]{#1}%
\providecommand \citenamefont [1]{#1}%
\providecommand \href@noop [0]{\@secondoftwo}%
\providecommand \href [0]{\begingroup \@sanitize@url \@href}%
\providecommand \@href[1]{\@@startlink{#1}\@@href}%
\providecommand \@@href[1]{\endgroup#1\@@endlink}%
\providecommand \@sanitize@url [0]{\catcode `\\12\catcode `\$12\catcode
  `\&12\catcode `\#12\catcode `\^12\catcode `\_12\catcode `\%12\relax}%
\providecommand \@@startlink[1]{}%
\providecommand \@@endlink[0]{}%
\providecommand \url  [0]{\begingroup\@sanitize@url \@url }%
\providecommand \@url [1]{\endgroup\@href {#1}{\urlprefix }}%
\providecommand \urlprefix  [0]{URL }%
\providecommand \Eprint [0]{\href }%
\providecommand \doibase [0]{https://doi.org/}%
\providecommand \selectlanguage [0]{\@gobble}%
\providecommand \bibinfo  [0]{\@secondoftwo}%
\providecommand \bibfield  [0]{\@secondoftwo}%
\providecommand \translation [1]{[#1]}%
\providecommand \BibitemOpen [0]{}%
\providecommand \bibitemStop [0]{}%
\providecommand \bibitemNoStop [0]{.\EOS\space}%
\providecommand \EOS [0]{\spacefactor3000\relax}%
\providecommand \BibitemShut  [1]{\csname bibitem#1\endcsname}%
\let\auto@bib@innerbib\@empty
\bibitem [{\citenamefont {Landau}(1933)}]{Landau1933}%
  \BibitemOpen
  \bibfield  {author} {\bibinfo {author} {\bibfnamefont {L.~D.}\ \bibnamefont
  {Landau}},\ }\href@noop {} {\bibfield  {journal} {\bibinfo  {journal} {Phys.
  Z. Sowjetunion}\ }\textbf {\bibinfo {volume} {3}},\ \bibinfo {pages} {644}
  (\bibinfo {year} {1933})}\BibitemShut {NoStop}%
\bibitem [{\citenamefont {Strinati}\ \emph {et~al.}(2018)\citenamefont
  {Strinati}, \citenamefont {Pieri}, \citenamefont {R{\"o}pke}, \citenamefont
  {Schuck},\ and\ \citenamefont {Urban}}]{Strinati2018tbb}%
  \BibitemOpen
  \bibfield  {author} {\bibinfo {author} {\bibfnamefont {G.~C.}\ \bibnamefont
  {Strinati}}, \bibinfo {author} {\bibfnamefont {P.}~\bibnamefont {Pieri}},
  \bibinfo {author} {\bibfnamefont {G.}~\bibnamefont {R{\"o}pke}}, \bibinfo
  {author} {\bibfnamefont {P.}~\bibnamefont {Schuck}},\ and\ \bibinfo {author}
  {\bibfnamefont {M.}~\bibnamefont {Urban}},\ }\href
  {https://doi.org/10.1016/j.physrep.2018.02.004} {\bibfield  {journal}
  {\bibinfo  {journal} {Phys. Rep.}\ }\textbf {\bibinfo {volume} {738}},\
  \bibinfo {pages} {1} (\bibinfo {year} {2018})}\BibitemShut {NoStop}%
\bibitem [{\citenamefont {Wölfle}(2018)}]{woelfle2018qic}%
  \BibitemOpen
  \bibfield  {author} {\bibinfo {author} {\bibfnamefont {P.}~\bibnamefont
  {Wölfle}},\ }\href
  {https://iopscience.iop.org/article/10.1088/1361-6633/aa9bc4/pdf} {\bibfield
  {journal} {\bibinfo  {journal} {Rep. Prog. Phys.}\ }\textbf {\bibinfo
  {volume} {81}},\ \bibinfo {pages} {032501} (\bibinfo {year}
  {2018})}\BibitemShut {NoStop}%
\bibitem [{\citenamefont {Bowley}(1973)}]{Bowley1973mto}%
  \BibitemOpen
  \bibfield  {author} {\bibinfo {author} {\bibfnamefont {R.~M.}\ \bibnamefont
  {Bowley}},\ }\href {https://doi.org/10.1007/BF00654922} {\bibfield  {journal}
  {\bibinfo  {journal} {J. Low Temp. Phys.}\ }\textbf {\bibinfo {volume}
  {10}},\ \bibinfo {pages} {481} (\bibinfo {year} {1973})}\BibitemShut
  {NoStop}%
\bibitem [{\citenamefont {Massignan}\ \emph {et~al.}(2014)\citenamefont
  {Massignan}, \citenamefont {Zaccanti},\ and\ \citenamefont
  {Bruun}}]{Massignan2014pdm}%
  \BibitemOpen
  \bibfield  {author} {\bibinfo {author} {\bibfnamefont {P.}~\bibnamefont
  {Massignan}}, \bibinfo {author} {\bibfnamefont {M.}~\bibnamefont
  {Zaccanti}},\ and\ \bibinfo {author} {\bibfnamefont {G.~M.}\ \bibnamefont
  {Bruun}},\ }\href {https://doi.org/10.1088/0034-4885/77/3/034401} {\bibfield
  {journal} {\bibinfo  {journal} {Rep. Prog. Phys.}\ }\textbf {\bibinfo
  {volume} {77}},\ \bibinfo {pages} {034401} (\bibinfo {year}
  {2014})}\BibitemShut {NoStop}%
\bibitem [{\citenamefont {Baym}\ and\ \citenamefont
  {Pethick}(1991)}]{Baym1991lfl}%
  \BibitemOpen
  \bibfield  {author} {\bibinfo {author} {\bibfnamefont {G.}~\bibnamefont
  {Baym}}\ and\ \bibinfo {author} {\bibfnamefont {C.}~\bibnamefont {Pethick}},\
  }\href {https://doi.org/10.1002/9783527617159} {\emph {\bibinfo {title}
  {Landau Fermi-Liquid Theory: Concepts and Applications}}}\ (\bibinfo
  {publisher} {Wiley-VCH},\ \bibinfo {year} {1991})\BibitemShut {NoStop}%
\bibitem [{\citenamefont {Norman}(2011)}]{Norman2011}%
  \BibitemOpen
  \bibfield  {author} {\bibinfo {author} {\bibfnamefont {M.~R.}\ \bibnamefont
  {Norman}},\ }\href {https://doi.org/10.1126/science.1200181} {\bibfield
  {journal} {\bibinfo  {journal} {Science}\ }\textbf {\bibinfo {volume}
  {332}},\ \bibinfo {pages} {196} (\bibinfo {year} {2011})}\BibitemShut
  {NoStop}%
\bibitem [{\citenamefont {Varma}\ \emph {et~al.}(2002)\citenamefont {Varma},
  \citenamefont {Nussinov},\ and\ \citenamefont {van Saarloos}}]{Varma2002}%
  \BibitemOpen
  \bibfield  {author} {\bibinfo {author} {\bibfnamefont {C.~M.}\ \bibnamefont
  {Varma}}, \bibinfo {author} {\bibfnamefont {Z.}~\bibnamefont {Nussinov}},\
  and\ \bibinfo {author} {\bibfnamefont {W.}~\bibnamefont {van Saarloos}},\
  }\href {https://doi.org/10.1016/S0370-1573(01)00060-6} {\bibfield  {journal}
  {\bibinfo  {journal} {Phys. Rep.}\ }\textbf {\bibinfo {volume} {361}},\
  \bibinfo {pages} {267} (\bibinfo {year} {2002})}\BibitemShut {NoStop}%
\bibitem [{\citenamefont {Bloch}\ \emph {et~al.}(2008)\citenamefont {Bloch},
  \citenamefont {Dalibard},\ and\ \citenamefont {Zwerger}}]{Bloch2008mbp}%
  \BibitemOpen
  \bibfield  {author} {\bibinfo {author} {\bibfnamefont {I.}~\bibnamefont
  {Bloch}}, \bibinfo {author} {\bibfnamefont {J.}~\bibnamefont {Dalibard}},\
  and\ \bibinfo {author} {\bibfnamefont {W.}~\bibnamefont {Zwerger}},\ }\href
  {https://doi.org/10.1103/RevModPhys.80.885} {\bibfield  {journal} {\bibinfo
  {journal} {Rev. Mod. Phys.}\ }\textbf {\bibinfo {volume} {80}},\ \bibinfo
  {pages} {885} (\bibinfo {year} {2008})}\BibitemShut {NoStop}%
\bibitem [{\citenamefont {Schirotzek}\ \emph {et~al.}(2009)\citenamefont
  {Schirotzek}, \citenamefont {Wu}, \citenamefont {Sommer},\ and\ \citenamefont
  {Zwierlein}}]{Schirotzek2009oof}%
  \BibitemOpen
  \bibfield  {author} {\bibinfo {author} {\bibfnamefont {A.}~\bibnamefont
  {Schirotzek}}, \bibinfo {author} {\bibfnamefont {C.-H.}\ \bibnamefont {Wu}},
  \bibinfo {author} {\bibfnamefont {A.}~\bibnamefont {Sommer}},\ and\ \bibinfo
  {author} {\bibfnamefont {M.~W.}\ \bibnamefont {Zwierlein}},\ }\href
  {https://doi.org/10.1103/PhysRevLett.102.230402} {\bibfield  {journal}
  {\bibinfo  {journal} {Phys. Rev. Lett.}\ }\textbf {\bibinfo {volume} {102}},\
  \bibinfo {pages} {230402} (\bibinfo {year} {2009})}\BibitemShut {NoStop}%
\bibitem [{\citenamefont {Nascimb\`ene}\ \emph {et~al.}(2009)\citenamefont
  {Nascimb\`ene}, \citenamefont {Navon}, \citenamefont {Jiang}, \citenamefont
  {Tarruell}, \citenamefont {Teichmann}, \citenamefont {McKeever},
  \citenamefont {Chevy},\ and\ \citenamefont {Salomon}}]{Nascimbene2009coo}%
  \BibitemOpen
  \bibfield  {author} {\bibinfo {author} {\bibfnamefont {S.}~\bibnamefont
  {Nascimb\`ene}}, \bibinfo {author} {\bibfnamefont {N.}~\bibnamefont {Navon}},
  \bibinfo {author} {\bibfnamefont {K.~J.}\ \bibnamefont {Jiang}}, \bibinfo
  {author} {\bibfnamefont {L.}~\bibnamefont {Tarruell}}, \bibinfo {author}
  {\bibfnamefont {M.}~\bibnamefont {Teichmann}}, \bibinfo {author}
  {\bibfnamefont {J.}~\bibnamefont {McKeever}}, \bibinfo {author}
  {\bibfnamefont {F.}~\bibnamefont {Chevy}},\ and\ \bibinfo {author}
  {\bibfnamefont {C.}~\bibnamefont {Salomon}},\ }\href
  {https://doi.org/10.1103/PhysRevLett.103.170402} {\bibfield  {journal}
  {\bibinfo  {journal} {Phys. Rev. Lett.}\ }\textbf {\bibinfo {volume} {103}},\
  \bibinfo {pages} {170402} (\bibinfo {year} {2009})}\BibitemShut {NoStop}%
\bibitem [{\citenamefont {Kohstall}\ \emph {et~al.}(2012)\citenamefont
  {Kohstall}, \citenamefont {Zaccanti}, \citenamefont {Jag}, \citenamefont
  {Trenkwalder}, \citenamefont {Massignan}, \citenamefont {Bruun},
  \citenamefont {Schreck},\ and\ \citenamefont {Grimm}}]{Kohstall2012mac}%
  \BibitemOpen
  \bibfield  {author} {\bibinfo {author} {\bibfnamefont {C.}~\bibnamefont
  {Kohstall}}, \bibinfo {author} {\bibfnamefont {M.}~\bibnamefont {Zaccanti}},
  \bibinfo {author} {\bibfnamefont {M.}~\bibnamefont {Jag}}, \bibinfo {author}
  {\bibfnamefont {A.}~\bibnamefont {Trenkwalder}}, \bibinfo {author}
  {\bibfnamefont {P.}~\bibnamefont {Massignan}}, \bibinfo {author}
  {\bibfnamefont {G.~M.}\ \bibnamefont {Bruun}}, \bibinfo {author}
  {\bibfnamefont {F.}~\bibnamefont {Schreck}},\ and\ \bibinfo {author}
  {\bibfnamefont {R.}~\bibnamefont {Grimm}},\ }\href
  {https://doi.org/10.1038/nature11065} {\bibfield  {journal} {\bibinfo
  {journal} {Nature (London)}\ }\textbf {\bibinfo {volume} {485}},\ \bibinfo
  {pages} {615} (\bibinfo {year} {2012})}\BibitemShut {NoStop}%
\bibitem [{\citenamefont {Cetina}\ \emph {et~al.}(2015)\citenamefont {Cetina},
  \citenamefont {Jag}, \citenamefont {Lous}, \citenamefont {Walraven},
  \citenamefont {Grimm}, \citenamefont {Christensen},\ and\ \citenamefont
  {Bruun}}]{Cetina2015doi}%
  \BibitemOpen
  \bibfield  {author} {\bibinfo {author} {\bibfnamefont {M.}~\bibnamefont
  {Cetina}}, \bibinfo {author} {\bibfnamefont {M.}~\bibnamefont {Jag}},
  \bibinfo {author} {\bibfnamefont {R.~S.}\ \bibnamefont {Lous}}, \bibinfo
  {author} {\bibfnamefont {J.~T.~M.}\ \bibnamefont {Walraven}}, \bibinfo
  {author} {\bibfnamefont {R.}~\bibnamefont {Grimm}}, \bibinfo {author}
  {\bibfnamefont {R.~S.}\ \bibnamefont {Christensen}},\ and\ \bibinfo {author}
  {\bibfnamefont {G.~M.}\ \bibnamefont {Bruun}},\ }\href
  {https://doi.org/10.1103/PhysRevLett.115.135302} {\bibfield  {journal}
  {\bibinfo  {journal} {Phys. Rev. Lett.}\ }\textbf {\bibinfo {volume} {115}},\
  \bibinfo {pages} {135302} (\bibinfo {year} {2015})}\BibitemShut {NoStop}%
\bibitem [{\citenamefont {Cetina}\ \emph {et~al.}(2016)\citenamefont {Cetina},
  \citenamefont {Jag}, \citenamefont {Lous}, \citenamefont {Fritsche},
  \citenamefont {Walraven}, \citenamefont {Grimm}, \citenamefont {Levinsen},
  \citenamefont {Parish}, \citenamefont {Schmidt}, \citenamefont {Knap},\ and\
  \citenamefont {Demler}}]{Cetina2016umb}%
  \BibitemOpen
  \bibfield  {author} {\bibinfo {author} {\bibfnamefont {M.}~\bibnamefont
  {Cetina}}, \bibinfo {author} {\bibfnamefont {M.}~\bibnamefont {Jag}},
  \bibinfo {author} {\bibfnamefont {R.~S.}\ \bibnamefont {Lous}}, \bibinfo
  {author} {\bibfnamefont {I.}~\bibnamefont {Fritsche}}, \bibinfo {author}
  {\bibfnamefont {J.~T.~M.}\ \bibnamefont {Walraven}}, \bibinfo {author}
  {\bibfnamefont {R.}~\bibnamefont {Grimm}}, \bibinfo {author} {\bibfnamefont
  {J.}~\bibnamefont {Levinsen}}, \bibinfo {author} {\bibfnamefont {M.~M.}\
  \bibnamefont {Parish}}, \bibinfo {author} {\bibfnamefont {R.}~\bibnamefont
  {Schmidt}}, \bibinfo {author} {\bibfnamefont {M.}~\bibnamefont {Knap}},\ and\
  \bibinfo {author} {\bibfnamefont {E.}~\bibnamefont {Demler}},\ }\href
  {https://doi.org/10.1126/science.aaf5134} {\bibfield  {journal} {\bibinfo
  {journal} {Science}\ }\textbf {\bibinfo {volume} {354}},\ \bibinfo {pages}
  {96} (\bibinfo {year} {2016})}\BibitemShut {NoStop}%
\bibitem [{\citenamefont {Scazza}\ \emph {et~al.}(2017)\citenamefont {Scazza},
  \citenamefont {Valtolina}, \citenamefont {Massignan}, \citenamefont {Recati},
  \citenamefont {Amico}, \citenamefont {Burchianti}, \citenamefont {Fort},
  \citenamefont {Inguscio}, \citenamefont {Zaccanti},\ and\ \citenamefont
  {Roati}}]{Scazza2017rfp}%
  \BibitemOpen
  \bibfield  {author} {\bibinfo {author} {\bibfnamefont {F.}~\bibnamefont
  {Scazza}}, \bibinfo {author} {\bibfnamefont {G.}~\bibnamefont {Valtolina}},
  \bibinfo {author} {\bibfnamefont {P.}~\bibnamefont {Massignan}}, \bibinfo
  {author} {\bibfnamefont {A.}~\bibnamefont {Recati}}, \bibinfo {author}
  {\bibfnamefont {A.}~\bibnamefont {Amico}}, \bibinfo {author} {\bibfnamefont
  {A.}~\bibnamefont {Burchianti}}, \bibinfo {author} {\bibfnamefont
  {C.}~\bibnamefont {Fort}}, \bibinfo {author} {\bibfnamefont {M.}~\bibnamefont
  {Inguscio}}, \bibinfo {author} {\bibfnamefont {M.}~\bibnamefont {Zaccanti}},\
  and\ \bibinfo {author} {\bibfnamefont {G.}~\bibnamefont {Roati}},\ }\href
  {https://doi.org/10.1103/PhysRevLett.118.083602} {\bibfield  {journal}
  {\bibinfo  {journal} {Phys. Rev. Lett.}\ }\textbf {\bibinfo {volume} {118}},\
  \bibinfo {pages} {083602} (\bibinfo {year} {2017})}\BibitemShut {NoStop}%
\bibitem [{\citenamefont {Darkwah~Oppong}\ \emph {et~al.}(2019)\citenamefont
  {Darkwah~Oppong}, \citenamefont {Riegger}, \citenamefont {Bettermann},
  \citenamefont {H{\"o}fer}, \citenamefont {Levinsen}, \citenamefont {Parish},
  \citenamefont {Bloch},\ and\ \citenamefont
  {F{\"o}lling}}]{DarkwahOppong2019ooc}%
  \BibitemOpen
  \bibfield  {author} {\bibinfo {author} {\bibfnamefont {N.}~\bibnamefont
  {Darkwah~Oppong}}, \bibinfo {author} {\bibfnamefont {L.}~\bibnamefont
  {Riegger}}, \bibinfo {author} {\bibfnamefont {O.}~\bibnamefont {Bettermann}},
  \bibinfo {author} {\bibfnamefont {M.}~\bibnamefont {H{\"o}fer}}, \bibinfo
  {author} {\bibfnamefont {J.}~\bibnamefont {Levinsen}}, \bibinfo {author}
  {\bibfnamefont {M.~M.}\ \bibnamefont {Parish}}, \bibinfo {author}
  {\bibfnamefont {I.}~\bibnamefont {Bloch}},\ and\ \bibinfo {author}
  {\bibfnamefont {S.}~\bibnamefont {F{\"o}lling}},\ }\href
  {https://doi.org/10.1103/PhysRevLett.122.193604} {\bibfield  {journal}
  {\bibinfo  {journal} {Physical Review Letters}\ }\textbf {\bibinfo {volume}
  {122}},\ \bibinfo {pages} {193604} (\bibinfo {year} {2019})}\BibitemShut
  {NoStop}%
\bibitem [{\citenamefont {Ness}\ \emph {et~al.}(2020)\citenamefont {Ness},
  \citenamefont {Shkedrov}, \citenamefont {Florshaim}, \citenamefont {Diessel},
  \citenamefont {von Milczewski}, \citenamefont {Schmidt},\ and\ \citenamefont
  {Sagi}}]{Ness2020ooa}%
  \BibitemOpen
  \bibfield  {author} {\bibinfo {author} {\bibfnamefont {G.}~\bibnamefont
  {Ness}}, \bibinfo {author} {\bibfnamefont {C.}~\bibnamefont {Shkedrov}},
  \bibinfo {author} {\bibfnamefont {Y.}~\bibnamefont {Florshaim}}, \bibinfo
  {author} {\bibfnamefont {O.~K.}\ \bibnamefont {Diessel}}, \bibinfo {author}
  {\bibfnamefont {J.}~\bibnamefont {von Milczewski}}, \bibinfo {author}
  {\bibfnamefont {R.}~\bibnamefont {Schmidt}},\ and\ \bibinfo {author}
  {\bibfnamefont {Y.}~\bibnamefont {Sagi}},\ }\href
  {https://doi.org/10.1103/PhysRevX.10.041019} {\bibfield  {journal} {\bibinfo
  {journal} {Phys. Rev. X}\ }\textbf {\bibinfo {volume} {10}},\ \bibinfo
  {pages} {041019} (\bibinfo {year} {2020})}\BibitemShut {NoStop}%
\bibitem [{\citenamefont {Adlong}\ \emph {et~al.}(2020)\citenamefont {Adlong},
  \citenamefont {Liu}, \citenamefont {Scazza}, \citenamefont {Zaccanti},
  \citenamefont {Oppong}, \citenamefont {F\"olling}, \citenamefont {Parish},\
  and\ \citenamefont {Levinsen}}]{Adlong2020}%
  \BibitemOpen
  \bibfield  {author} {\bibinfo {author} {\bibfnamefont {H.~S.}\ \bibnamefont
  {Adlong}}, \bibinfo {author} {\bibfnamefont {W.~E.}\ \bibnamefont {Liu}},
  \bibinfo {author} {\bibfnamefont {F.}~\bibnamefont {Scazza}}, \bibinfo
  {author} {\bibfnamefont {M.}~\bibnamefont {Zaccanti}}, \bibinfo {author}
  {\bibfnamefont {N.~D.}\ \bibnamefont {Oppong}}, \bibinfo {author}
  {\bibfnamefont {S.}~\bibnamefont {F\"olling}}, \bibinfo {author}
  {\bibfnamefont {M.~M.}\ \bibnamefont {Parish}},\ and\ \bibinfo {author}
  {\bibfnamefont {J.}~\bibnamefont {Levinsen}},\ }\href
  {https://doi.org/https://doi.org/10.1103/PhysRevLett.125.133401} {\bibfield
  {journal} {\bibinfo  {journal} {Phys. Rev. Lett.}\ }\textbf {\bibinfo
  {volume} {125}},\ \bibinfo {pages} {133401} (\bibinfo {year}
  {2020})}\BibitemShut {NoStop}%
\bibitem [{\citenamefont {Hu}\ \emph {et~al.}(2016)\citenamefont {Hu},
  \citenamefont {Van~de Graaff}, \citenamefont {Kedar}, \citenamefont {Corson},
  \citenamefont {Cornell},\ and\ \citenamefont {Jin}}]{Hu2016bpi}%
  \BibitemOpen
  \bibfield  {author} {\bibinfo {author} {\bibfnamefont {M.-G.}\ \bibnamefont
  {Hu}}, \bibinfo {author} {\bibfnamefont {M.~J.}\ \bibnamefont {Van~de
  Graaff}}, \bibinfo {author} {\bibfnamefont {D.}~\bibnamefont {Kedar}},
  \bibinfo {author} {\bibfnamefont {J.~P.}\ \bibnamefont {Corson}}, \bibinfo
  {author} {\bibfnamefont {E.~A.}\ \bibnamefont {Cornell}},\ and\ \bibinfo
  {author} {\bibfnamefont {D.~S.}\ \bibnamefont {Jin}},\ }\href
  {https://doi.org/10.1103/PhysRevLett.117.055301} {\bibfield  {journal}
  {\bibinfo  {journal} {Phys. Rev. Lett.}\ }\textbf {\bibinfo {volume} {117}},\
  \bibinfo {pages} {055301} (\bibinfo {year} {2016})}\BibitemShut {NoStop}%
\bibitem [{\citenamefont {J{\o}rgensen}\ \emph {et~al.}(2016)\citenamefont
  {J{\o}rgensen}, \citenamefont {Wacker}, \citenamefont {Skalmstang},
  \citenamefont {Parish}, \citenamefont {Levinsen}, \citenamefont
  {Christensen}, \citenamefont {Bruun},\ and\ \citenamefont
  {Arlt}}]{Jorgensen2016ooa}%
  \BibitemOpen
  \bibfield  {author} {\bibinfo {author} {\bibfnamefont {N.~B.}\ \bibnamefont
  {J{\o}rgensen}}, \bibinfo {author} {\bibfnamefont {L.}~\bibnamefont
  {Wacker}}, \bibinfo {author} {\bibfnamefont {K.~T.}\ \bibnamefont
  {Skalmstang}}, \bibinfo {author} {\bibfnamefont {M.~M.}\ \bibnamefont
  {Parish}}, \bibinfo {author} {\bibfnamefont {J.}~\bibnamefont {Levinsen}},
  \bibinfo {author} {\bibfnamefont {R.~S.}\ \bibnamefont {Christensen}},
  \bibinfo {author} {\bibfnamefont {G.~M.}\ \bibnamefont {Bruun}},\ and\
  \bibinfo {author} {\bibfnamefont {J.~J.}\ \bibnamefont {Arlt}},\ }\href
  {https://doi.org/10.1103/PhysRevLett.117.055302} {\bibfield  {journal}
  {\bibinfo  {journal} {Phys. Rev. Lett.}\ }\textbf {\bibinfo {volume} {117}},\
  \bibinfo {pages} {055302} (\bibinfo {year} {2016})}\BibitemShut {NoStop}%
\bibitem [{\citenamefont {Pe\~na Ardila}\ \emph {et~al.}(2019)\citenamefont
  {Pe\~na Ardila}, \citenamefont {J\o{}rgensen}, \citenamefont {Pohl},
  \citenamefont {Giorgini}, \citenamefont {Bruun},\ and\ \citenamefont
  {Arlt}}]{Ardila2019}%
  \BibitemOpen
  \bibfield  {author} {\bibinfo {author} {\bibfnamefont {L.~A.}\ \bibnamefont
  {Pe\~na Ardila}}, \bibinfo {author} {\bibfnamefont {N.~B.}\ \bibnamefont
  {J\o{}rgensen}}, \bibinfo {author} {\bibfnamefont {T.}~\bibnamefont {Pohl}},
  \bibinfo {author} {\bibfnamefont {S.}~\bibnamefont {Giorgini}}, \bibinfo
  {author} {\bibfnamefont {G.~M.}\ \bibnamefont {Bruun}},\ and\ \bibinfo
  {author} {\bibfnamefont {J.~J.}\ \bibnamefont {Arlt}},\ }\href
  {https://doi.org/10.1103/PhysRevA.99.063607} {\bibfield  {journal} {\bibinfo
  {journal} {Phys. Rev. A}\ }\textbf {\bibinfo {volume} {99}},\ \bibinfo
  {pages} {063607} (\bibinfo {year} {2019})}\BibitemShut {NoStop}%
\bibitem [{\citenamefont {Yan}\ \emph {et~al.}(2020)\citenamefont {Yan},
  \citenamefont {Ni}, \citenamefont {Robens},\ and\ \citenamefont
  {Zwierlein}}]{Yan2020bpn}%
  \BibitemOpen
  \bibfield  {author} {\bibinfo {author} {\bibfnamefont {Z.~Z.}\ \bibnamefont
  {Yan}}, \bibinfo {author} {\bibfnamefont {Y.}~\bibnamefont {Ni}}, \bibinfo
  {author} {\bibfnamefont {C.}~\bibnamefont {Robens}},\ and\ \bibinfo {author}
  {\bibfnamefont {M.~W.}\ \bibnamefont {Zwierlein}},\ }\href
  {https://doi.org/10.1126/science.aax5850} {\bibfield  {journal} {\bibinfo
  {journal} {Science}\ }\textbf {\bibinfo {volume} {368}},\ \bibinfo {pages}
  {6487} (\bibinfo {year} {2020})}\BibitemShut {NoStop}%
\bibitem [{\citenamefont {Skou}\ \emph {et~al.}(2020)\citenamefont {Skou},
  \citenamefont {Skov}, \citenamefont {J{\o}rgensen}, \citenamefont {Nielsen},
  \citenamefont {Camacho-Guardian}, \citenamefont {Pohl}, \citenamefont
  {Bruun},\ and\ \citenamefont {Arlt}}]{Skou2020}%
  \BibitemOpen
  \bibfield  {author} {\bibinfo {author} {\bibfnamefont {M.~G.}\ \bibnamefont
  {Skou}}, \bibinfo {author} {\bibfnamefont {T.~G.}\ \bibnamefont {Skov}},
  \bibinfo {author} {\bibfnamefont {N.~B.}\ \bibnamefont {J{\o}rgensen}},
  \bibinfo {author} {\bibfnamefont {K.~K.}\ \bibnamefont {Nielsen}}, \bibinfo
  {author} {\bibfnamefont {A.}~\bibnamefont {Camacho-Guardian}}, \bibinfo
  {author} {\bibfnamefont {T.}~\bibnamefont {Pohl}}, \bibinfo {author}
  {\bibfnamefont {G.~M.}\ \bibnamefont {Bruun}},\ and\ \bibinfo {author}
  {\bibfnamefont {J.~J.}\ \bibnamefont {Arlt}},\ }\href
  {https://arxiv.org/abs/2005.00424v1} {\bibfield  {journal} {\bibinfo
  {journal} {arXiv}\ } (\bibinfo {year} {2020})},\ \Eprint
  {https://arxiv.org/abs/2005.00424} {2005.00424} \BibitemShut {NoStop}%
\bibitem [{\citenamefont {Chevy}\ and\ \citenamefont
  {Mora}(2010)}]{Chevy2010ucp}%
  \BibitemOpen
  \bibfield  {author} {\bibinfo {author} {\bibfnamefont {F.}~\bibnamefont
  {Chevy}}\ and\ \bibinfo {author} {\bibfnamefont {C.}~\bibnamefont {Mora}},\
  }\href {https://doi.org/10.1088/0034-4885/73/11/112401} {\bibfield  {journal}
  {\bibinfo  {journal} {Rep. Prog. Phys.}\ }\textbf {\bibinfo {volume} {73}},\
  \bibinfo {pages} {112401} (\bibinfo {year} {2010})}\BibitemShut {NoStop}%
\bibitem [{\citenamefont {Levinsen}\ and\ \citenamefont
  {Parish}(2015)}]{Levinsen2015Feb}%
  \BibitemOpen
  \bibfield  {author} {\bibinfo {author} {\bibfnamefont {J.}~\bibnamefont
  {Levinsen}}\ and\ \bibinfo {author} {\bibfnamefont {M.~M.}\ \bibnamefont
  {Parish}},\ }in\ \href {https://doi.org/10.1142/9789814667746_0001} {\emph
  {\bibinfo {booktitle} {{Annual Review of Cold Atoms and Molecules}}}},\
  Vol.~\bibinfo {volume} {3}\ (\bibinfo  {publisher} {World Scientific},\
  \bibinfo {address} {Singapore},\ \bibinfo {year} {2015})\ pp.\ \bibinfo
  {pages} {1--75}\BibitemShut {NoStop}%
\bibitem [{\citenamefont {Schmidt}\ \emph {et~al.}(2018)\citenamefont
  {Schmidt}, \citenamefont {Knap}, \citenamefont {Ivanov}, \citenamefont {You},
  \citenamefont {Cetina},\ and\ \citenamefont {Demler}}]{Schmidt2018umb}%
  \BibitemOpen
  \bibfield  {author} {\bibinfo {author} {\bibfnamefont {R.}~\bibnamefont
  {Schmidt}}, \bibinfo {author} {\bibfnamefont {M.}~\bibnamefont {Knap}},
  \bibinfo {author} {\bibfnamefont {D.~A.}\ \bibnamefont {Ivanov}}, \bibinfo
  {author} {\bibfnamefont {J.-S.}\ \bibnamefont {You}}, \bibinfo {author}
  {\bibfnamefont {M.}~\bibnamefont {Cetina}},\ and\ \bibinfo {author}
  {\bibfnamefont {E.}~\bibnamefont {Demler}},\ }\href
  {https://doi.org/10.1088/1361-6633/aa9593} {\bibfield  {journal} {\bibinfo
  {journal} {Rep. Prog. Phys.}\ }\textbf {\bibinfo {volume} {81}},\ \bibinfo
  {pages} {024401} (\bibinfo {year} {2018})}\BibitemShut {NoStop}%
\bibitem [{\citenamefont {Pitaevskii}\ and\ \citenamefont
  {Stringari}(2016)}]{Pitaevskii2016book}%
  \BibitemOpen
  \bibfield  {author} {\bibinfo {author} {\bibfnamefont {L.}~\bibnamefont
  {Pitaevskii}}\ and\ \bibinfo {author} {\bibfnamefont {S.}~\bibnamefont
  {Stringari}},\ }\href
  {https://doi.org/10.1093/acprof:oso/9780198758884.001.0001} {\emph {\bibinfo
  {title} {Bose-Einstein Condensation and Superfluidity}}}\ (\bibinfo
  {publisher} {Oxford University Press},\ \bibinfo {year} {2016})\BibitemShut
  {NoStop}%
\bibitem [{\citenamefont {Kinnunen}\ \emph {et~al.}(2018)\citenamefont
  {Kinnunen}, \citenamefont {Baarsma}, \citenamefont {Martikainen},\ and\
  \citenamefont {Törmä}}]{Kinnunen2018}%
  \BibitemOpen
  \bibfield  {author} {\bibinfo {author} {\bibfnamefont {J.~J.}\ \bibnamefont
  {Kinnunen}}, \bibinfo {author} {\bibfnamefont {J.~E.}\ \bibnamefont
  {Baarsma}}, \bibinfo {author} {\bibfnamefont {J.-P.}\ \bibnamefont
  {Martikainen}},\ and\ \bibinfo {author} {\bibfnamefont {P.}~\bibnamefont
  {Törmä}},\ }\href {https://doi.org/10.1088/1361-6633/aaa4ad} {\bibfield
  {journal} {\bibinfo  {journal} {Reports on Progress in Physics}\ }\textbf
  {\bibinfo {volume} {81}},\ \bibinfo {pages} {046401} (\bibinfo {year}
  {2018})}\BibitemShut {NoStop}%
\bibitem [{\citenamefont {Lous}\ \emph {et~al.}(2017)\citenamefont {Lous},
  \citenamefont {Fritsche}, \citenamefont {Jag}, \citenamefont {Huang},\ and\
  \citenamefont {Grimm}}]{Lous2017toa}%
  \BibitemOpen
  \bibfield  {author} {\bibinfo {author} {\bibfnamefont {R.~S.}\ \bibnamefont
  {Lous}}, \bibinfo {author} {\bibfnamefont {I.}~\bibnamefont {Fritsche}},
  \bibinfo {author} {\bibfnamefont {M.}~\bibnamefont {Jag}}, \bibinfo {author}
  {\bibfnamefont {B.}~\bibnamefont {Huang}},\ and\ \bibinfo {author}
  {\bibfnamefont {R.}~\bibnamefont {Grimm}},\ }\href
  {https://doi.org/10.1103/PhysRevA.95.053627} {\bibfield  {journal} {\bibinfo
  {journal} {Phys. Rev. A}\ }\textbf {\bibinfo {volume} {95}},\ \bibinfo
  {pages} {053627} (\bibinfo {year} {2017})}\BibitemShut {NoStop}%
\bibitem [{\citenamefont {Lous}\ \emph {et~al.}(2018)\citenamefont {Lous},
  \citenamefont {Fritsche}, \citenamefont {Jag}, \citenamefont {Lehmann},
  \citenamefont {Kirilov}, \citenamefont {Huang},\ and\ \citenamefont
  {Grimm}}]{Lous2018pti}%
  \BibitemOpen
  \bibfield  {author} {\bibinfo {author} {\bibfnamefont {R.~S.}\ \bibnamefont
  {Lous}}, \bibinfo {author} {\bibfnamefont {I.}~\bibnamefont {Fritsche}},
  \bibinfo {author} {\bibfnamefont {M.}~\bibnamefont {Jag}}, \bibinfo {author}
  {\bibfnamefont {F.}~\bibnamefont {Lehmann}}, \bibinfo {author} {\bibfnamefont
  {E.}~\bibnamefont {Kirilov}}, \bibinfo {author} {\bibfnamefont
  {B.}~\bibnamefont {Huang}},\ and\ \bibinfo {author} {\bibfnamefont
  {R.}~\bibnamefont {Grimm}},\ }\href
  {https://doi.org/10.1103/PhysRevLett.120.243403} {\bibfield  {journal}
  {\bibinfo  {journal} {Phys. Rev. Lett.}\ }\textbf {\bibinfo {volume} {120}},\
  \bibinfo {pages} {243403} (\bibinfo {year} {2018})}\BibitemShut {NoStop}%
\bibitem [{\citenamefont {Huang}\ \emph {et~al.}(2019)\citenamefont {Huang},
  \citenamefont {Fritsche}, \citenamefont {Lous}, \citenamefont {Baroni},
  \citenamefont {Walraven}, \citenamefont {Kirilov},\ and\ \citenamefont
  {Grimm}}]{Huang2019}%
  \BibitemOpen
  \bibfield  {author} {\bibinfo {author} {\bibfnamefont {B.}~\bibnamefont
  {Huang}}, \bibinfo {author} {\bibfnamefont {I.}~\bibnamefont {Fritsche}},
  \bibinfo {author} {\bibfnamefont {R.~S.}\ \bibnamefont {Lous}}, \bibinfo
  {author} {\bibfnamefont {C.}~\bibnamefont {Baroni}}, \bibinfo {author}
  {\bibfnamefont {J.~T.~M.}\ \bibnamefont {Walraven}}, \bibinfo {author}
  {\bibfnamefont {E.}~\bibnamefont {Kirilov}},\ and\ \bibinfo {author}
  {\bibfnamefont {R.}~\bibnamefont {Grimm}},\ }\href
  {https://doi.org/https://doi.org/10.1103/PhysRevA.99.041602} {\bibfield
  {journal} {\bibinfo  {journal} {Phys. Rev. A}\ }\textbf {\bibinfo {volume}
  {99}},\ \bibinfo {pages} {041602(R)} (\bibinfo {year} {2019})}\BibitemShut
  {NoStop}%
\bibitem [{\citenamefont {Mora}\ and\ \citenamefont {Chevy}(2010)}]{Mora2010}%
  \BibitemOpen
  \bibfield  {author} {\bibinfo {author} {\bibfnamefont {C.}~\bibnamefont
  {Mora}}\ and\ \bibinfo {author} {\bibfnamefont {F.}~\bibnamefont {Chevy}},\
  }\href {https://doi.org/10.1103/PhysRevLett.104.230402} {\bibfield  {journal}
  {\bibinfo  {journal} {Phys. Rev. Lett.}\ }\textbf {\bibinfo {volume} {104}},\
  \bibinfo {pages} {230402} (\bibinfo {year} {2010})}\BibitemShut {NoStop}%
\bibitem [{\citenamefont {Yu}\ \emph {et~al.}(2010)\citenamefont {Yu},
  \citenamefont {Z\"ollner},\ and\ \citenamefont {Pethick}}]{Yu2010}%
  \BibitemOpen
  \bibfield  {author} {\bibinfo {author} {\bibfnamefont {Z.}~\bibnamefont
  {Yu}}, \bibinfo {author} {\bibfnamefont {S.}~\bibnamefont {Z\"ollner}},\ and\
  \bibinfo {author} {\bibfnamefont {C.~J.}\ \bibnamefont {Pethick}},\ }\href
  {https://doi.org/10.1103/PhysRevLett.105.188901} {\bibfield  {journal}
  {\bibinfo  {journal} {Phys. Rev. Lett.}\ }\textbf {\bibinfo {volume} {105}},\
  \bibinfo {pages} {188901} (\bibinfo {year} {2010})}\BibitemShut {NoStop}%
\bibitem [{\citenamefont {Yu}\ and\ \citenamefont {Pethick}(2012)}]{Yu2012}%
  \BibitemOpen
  \bibfield  {author} {\bibinfo {author} {\bibfnamefont {Z.}~\bibnamefont
  {Yu}}\ and\ \bibinfo {author} {\bibfnamefont {C.~J.}\ \bibnamefont
  {Pethick}},\ }\href {https://doi.org/10.1103/PhysRevA.85.063616} {\bibfield
  {journal} {\bibinfo  {journal} {Phys. Rev. A}\ }\textbf {\bibinfo {volume}
  {85}},\ \bibinfo {pages} {063616} (\bibinfo {year} {2012})}\BibitemShut
  {NoStop}%
\bibitem [{\citenamefont {Camacho-Guardian}\ and\ \citenamefont
  {Bruun}(2018)}]{CamachoLandau}%
  \BibitemOpen
  \bibfield  {author} {\bibinfo {author} {\bibfnamefont {A.}~\bibnamefont
  {Camacho-Guardian}}\ and\ \bibinfo {author} {\bibfnamefont {G.~M.}\
  \bibnamefont {Bruun}},\ }\href {https://doi.org/10.1103/PhysRevX.8.031042}
  {\bibfield  {journal} {\bibinfo  {journal} {Phys. Rev. X}\ }\textbf {\bibinfo
  {volume} {8}},\ \bibinfo {pages} {031042} (\bibinfo {year}
  {2018})}\BibitemShut {NoStop}%
\bibitem [{\citenamefont {Camacho-Guardian}\ \emph {et~al.}(2018)\citenamefont
  {Camacho-Guardian}, \citenamefont {Pe\~na Ardila}, \citenamefont {Pohl},\
  and\ \citenamefont {Bruun}}]{Camacho2018bia}%
  \BibitemOpen
  \bibfield  {author} {\bibinfo {author} {\bibfnamefont {A.}~\bibnamefont
  {Camacho-Guardian}}, \bibinfo {author} {\bibfnamefont {L.~A.}\ \bibnamefont
  {Pe\~na Ardila}}, \bibinfo {author} {\bibfnamefont {T.}~\bibnamefont
  {Pohl}},\ and\ \bibinfo {author} {\bibfnamefont {G.~M.}\ \bibnamefont
  {Bruun}},\ }\href
  {https://journals.aps.org/prl/abstract/10.1103/PhysRevLett.121.013401}
  {\bibfield  {journal} {\bibinfo  {journal} {Phys. Rev. Lett.}\ }\textbf
  {\bibinfo {volume} {121}},\ \bibinfo {pages} {013401} (\bibinfo {year}
  {2018})}\BibitemShut {NoStop}%
\bibitem [{\citenamefont {Chevy}(2006)}]{Chevy2006upd}%
  \BibitemOpen
  \bibfield  {author} {\bibinfo {author} {\bibfnamefont {F.}~\bibnamefont
  {Chevy}},\ }\href {https://doi.org/10.1103/PhysRevA.74.063628} {\bibfield
  {journal} {\bibinfo  {journal} {Phys. Rev. A}\ }\textbf {\bibinfo {volume}
  {74}},\ \bibinfo {pages} {063628} (\bibinfo {year} {2006})}\BibitemShut
  {NoStop}%
\bibitem [{\citenamefont {Massignan}(2012)}]{Massignan2012pad}%
  \BibitemOpen
  \bibfield  {author} {\bibinfo {author} {\bibfnamefont {P.}~\bibnamefont
  {Massignan}},\ }\href {https://doi.org/10.1209/0295-5075/98/10012} {\bibfield
   {journal} {\bibinfo  {journal} {Europhys. Lett.}\ }\textbf {\bibinfo
  {volume} {98}},\ \bibinfo {pages} {10012} (\bibinfo {year}
  {2012})}\BibitemShut {NoStop}%
\bibitem [{\citenamefont {Parish}\ and\ \citenamefont
  {Levinsen}(2016)}]{Parish2016qdo}%
  \BibitemOpen
  \bibfield  {author} {\bibinfo {author} {\bibfnamefont {M.~M.}\ \bibnamefont
  {Parish}}\ and\ \bibinfo {author} {\bibfnamefont {J.}~\bibnamefont
  {Levinsen}},\ }\href
  {https://journals.aps.org/prb/pdf/10.1103/PhysRevB.94.184303} {\bibfield
  {journal} {\bibinfo  {journal} {Phys. Rev. B}\ }\textbf {\bibinfo {volume}
  {94}},\ \bibinfo {pages} {184303} (\bibinfo {year} {2016})}\BibitemShut
  {NoStop}%
\bibitem [{\citenamefont {Santamore}\ and\ \citenamefont
  {Timmermans}(2008)}]{Santamore2008fmi}%
  \BibitemOpen
  \bibfield  {author} {\bibinfo {author} {\bibfnamefont {D.~H.}\ \bibnamefont
  {Santamore}}\ and\ \bibinfo {author} {\bibfnamefont {E.}~\bibnamefont
  {Timmermans}},\ }\href
  {https://journals.aps.org/pra/pdf/10.1103/PhysRevA.78.013619} {\bibfield
  {journal} {\bibinfo  {journal} {Phys. Rev. A}\ }\textbf {\bibinfo {volume}
  {78}},\ \bibinfo {pages} {013619} (\bibinfo {year} {2008})}\BibitemShut
  {NoStop}%
\bibitem [{\citenamefont {Hu}\ \emph {et~al.}(2018)\citenamefont {Hu},
  \citenamefont {Mulkerin}, \citenamefont {Wang},\ and\ \citenamefont
  {Liu}}]{Hu2018afp}%
  \BibitemOpen
  \bibfield  {author} {\bibinfo {author} {\bibfnamefont {H.}~\bibnamefont
  {Hu}}, \bibinfo {author} {\bibfnamefont {B.~C.}\ \bibnamefont {Mulkerin}},
  \bibinfo {author} {\bibfnamefont {J.}~\bibnamefont {Wang}},\ and\ \bibinfo
  {author} {\bibfnamefont {X.~J.}\ \bibnamefont {Liu}},\ }\href
  {https://journals.aps.org/pra/abstract/10.1103/PhysRevA.98.013626} {\bibfield
   {journal} {\bibinfo  {journal} {Phys. Rev. A}\ }\textbf {\bibinfo {volume}
  {98}},\ \bibinfo {pages} {013626} (\bibinfo {year} {2018})}\BibitemShut
  {NoStop}%
\bibitem [{\citenamefont {Tajima}\ and\ \citenamefont
  {Uchino}(2018)}]{Tajima2018mfp}%
  \BibitemOpen
  \bibfield  {author} {\bibinfo {author} {\bibfnamefont {H.}~\bibnamefont
  {Tajima}}\ and\ \bibinfo {author} {\bibfnamefont {S.}~\bibnamefont
  {Uchino}},\ }\href
  {https://iopscience.iop.org/article/10.1088/1367-2630/aad1e7/pdf} {\bibfield
  {journal} {\bibinfo  {journal} {New J. Phys.}\ }\textbf {\bibinfo {volume}
  {20}},\ \bibinfo {pages} {073048} (\bibinfo {year} {2018})}\BibitemShut
  {NoStop}%
\bibitem [{\citenamefont {DeSalvo}\ \emph {et~al.}(2019)\citenamefont
  {DeSalvo}, \citenamefont {Patel}, \citenamefont {Cai},\ and\ \citenamefont
  {Cheng}}]{DeSalvo2019oof}%
  \BibitemOpen
  \bibfield  {author} {\bibinfo {author} {\bibfnamefont {B.}~\bibnamefont
  {DeSalvo}}, \bibinfo {author} {\bibfnamefont {K.}~\bibnamefont {Patel}},
  \bibinfo {author} {\bibfnamefont {G.}~\bibnamefont {Cai}},\ and\ \bibinfo
  {author} {\bibfnamefont {C.}~\bibnamefont {Cheng}},\ }\href
  {https://doi.org/10.1038/s41586-019-1055-0} {\bibfield  {journal} {\bibinfo
  {journal} {Nature}\ }\textbf {\bibinfo {volume} {568}},\ \bibinfo {pages}
  {61} (\bibinfo {year} {2019})}\BibitemShut {NoStop}%
\bibitem [{\citenamefont {Edri}\ \emph {et~al.}(2020)\citenamefont {Edri},
  \citenamefont {Raz}, \citenamefont {Matzliah}, \citenamefont {Davidson},\
  and\ \citenamefont {Ozeri}}]{Edri2020oos}%
  \BibitemOpen
  \bibfield  {author} {\bibinfo {author} {\bibfnamefont {H.}~\bibnamefont
  {Edri}}, \bibinfo {author} {\bibfnamefont {B.}~\bibnamefont {Raz}}, \bibinfo
  {author} {\bibfnamefont {N.}~\bibnamefont {Matzliah}}, \bibinfo {author}
  {\bibfnamefont {N.}~\bibnamefont {Davidson}},\ and\ \bibinfo {author}
  {\bibfnamefont {R.}~\bibnamefont {Ozeri}},\ }\href
  {https://journals.aps.org/prl/abstract/10.1103/PhysRevLett.124.163401}
  {\bibfield  {journal} {\bibinfo  {journal} {Phys. Rev. Lett.}\ }\textbf
  {\bibinfo {volume} {124}},\ \bibinfo {pages} {163401} (\bibinfo {year}
  {2020})}\BibitemShut {NoStop}%
\bibitem [{\citenamefont {Chin}\ \emph {et~al.}(2010)\citenamefont {Chin},
  \citenamefont {Grimm}, \citenamefont {Julienne},\ and\ \citenamefont
  {Tiesinga}}]{Chin2010fri}%
  \BibitemOpen
  \bibfield  {author} {\bibinfo {author} {\bibfnamefont {C.}~\bibnamefont
  {Chin}}, \bibinfo {author} {\bibfnamefont {R.}~\bibnamefont {Grimm}},
  \bibinfo {author} {\bibfnamefont {P.~S.}\ \bibnamefont {Julienne}},\ and\
  \bibinfo {author} {\bibfnamefont {E.}~\bibnamefont {Tiesinga}},\ }\href
  {https://doi.org/10.1103/RevModPhys.82.1225} {\bibfield  {journal} {\bibinfo
  {journal} {Rev. Mod. Phys.}\ }\textbf {\bibinfo {volume} {82}},\ \bibinfo
  {pages} {1225} (\bibinfo {year} {2010})}\BibitemShut {NoStop}%
\bibitem [{\citenamefont {Naik}\ \emph {et~al.}(2011)\citenamefont {Naik},
  \citenamefont {Trenkwalder}, \citenamefont {Kohstall}, \citenamefont
  {Spiegelhalder}, \citenamefont {Zaccanti}, \citenamefont {Hendl},
  \citenamefont {Schreck}, \citenamefont {Grimm}, \citenamefont {Hanna},\ and\
  \citenamefont {Julienne}}]{Naik2011fri}%
  \BibitemOpen
  \bibfield  {author} {\bibinfo {author} {\bibfnamefont {D.}~\bibnamefont
  {Naik}}, \bibinfo {author} {\bibfnamefont {A.}~\bibnamefont {Trenkwalder}},
  \bibinfo {author} {\bibfnamefont {C.}~\bibnamefont {Kohstall}}, \bibinfo
  {author} {\bibfnamefont {F.~M.}\ \bibnamefont {Spiegelhalder}}, \bibinfo
  {author} {\bibfnamefont {M.}~\bibnamefont {Zaccanti}}, \bibinfo {author}
  {\bibfnamefont {G.}~\bibnamefont {Hendl}}, \bibinfo {author} {\bibfnamefont
  {F.}~\bibnamefont {Schreck}}, \bibinfo {author} {\bibfnamefont
  {R.}~\bibnamefont {Grimm}}, \bibinfo {author} {\bibfnamefont
  {T.}~\bibnamefont {Hanna}},\ and\ \bibinfo {author} {\bibfnamefont
  {P.}~\bibnamefont {Julienne}},\ }\href
  {https://doi.org/10.1140/epjd/e2010-10591-2} {\bibfield  {journal} {\bibinfo
  {journal} {Eur. Phys. J. D}\ }\textbf {\bibinfo {volume} {65}},\ \bibinfo
  {pages} {55} (\bibinfo {year} {2011})}\BibitemShut {NoStop}%
\bibitem [{\citenamefont {Jag}\ \emph {et~al.}(2016)\citenamefont {Jag},
  \citenamefont {Cetina}, \citenamefont {Lous}, \citenamefont {Grimm},
  \citenamefont {Levinsen},\ and\ \citenamefont {Petrov}}]{Jag2016lof}%
  \BibitemOpen
  \bibfield  {author} {\bibinfo {author} {\bibfnamefont {M.}~\bibnamefont
  {Jag}}, \bibinfo {author} {\bibfnamefont {M.}~\bibnamefont {Cetina}},
  \bibinfo {author} {\bibfnamefont {R.~S.}\ \bibnamefont {Lous}}, \bibinfo
  {author} {\bibfnamefont {R.}~\bibnamefont {Grimm}}, \bibinfo {author}
  {\bibfnamefont {J.}~\bibnamefont {Levinsen}},\ and\ \bibinfo {author}
  {\bibfnamefont {D.~S.}\ \bibnamefont {Petrov}},\ }\href
  {https://doi.org/10.1103/PhysRevA.94.062706} {\bibfield  {journal} {\bibinfo
  {journal} {Phys. Rev. A}\ }\textbf {\bibinfo {volume} {94}},\ \bibinfo
  {pages} {062706} (\bibinfo {year} {2016})}\BibitemShut {NoStop}%
\bibitem [{\citenamefont {Naidon}\ and\ \citenamefont
  {Endo}(2017)}]{Naidon2017epa}%
  \BibitemOpen
  \bibfield  {author} {\bibinfo {author} {\bibfnamefont {P.}~\bibnamefont
  {Naidon}}\ and\ \bibinfo {author} {\bibfnamefont {S.}~\bibnamefont {Endo}},\
  }\href {https://doi.org/10.1088/1361-6633/aa50e8} {\bibfield  {journal}
  {\bibinfo  {journal} {Rep. Prog. Phys.}\ }\textbf {\bibinfo {volume} {80}},\
  \bibinfo {pages} {056001} (\bibinfo {year} {2017})}\BibitemShut {NoStop}%
\bibitem [{\citenamefont {Greene}\ \emph {et~al.}(2017)\citenamefont {Greene},
  \citenamefont {Giannakeas},\ and\ \citenamefont
  {P\'erez-R\'{\i}os}}]{Greene2017ufb}%
  \BibitemOpen
  \bibfield  {author} {\bibinfo {author} {\bibfnamefont {C.~H.}\ \bibnamefont
  {Greene}}, \bibinfo {author} {\bibfnamefont {P.}~\bibnamefont {Giannakeas}},\
  and\ \bibinfo {author} {\bibfnamefont {J.}~\bibnamefont
  {P\'erez-R\'{\i}os}},\ }\href {https://doi.org/10.1103/RevModPhys.89.035006}
  {\bibfield  {journal} {\bibinfo  {journal} {Rev. Mod. Phys.}\ }\textbf
  {\bibinfo {volume} {89}},\ \bibinfo {pages} {035006} (\bibinfo {year}
  {2017})}\BibitemShut {NoStop}%
\bibitem [{\citenamefont {H\"afner}\ \emph {et~al.}(2017)\citenamefont
  {H\"afner}, \citenamefont {Ulmanis}, \citenamefont {Kuhnle}, \citenamefont
  {Wang}, \citenamefont {Greene},\ and\ \citenamefont
  {Weidem\"uller}}]{Haefner2017rot}%
  \BibitemOpen
  \bibfield  {author} {\bibinfo {author} {\bibfnamefont {S.}~\bibnamefont
  {H\"afner}}, \bibinfo {author} {\bibfnamefont {J.}~\bibnamefont {Ulmanis}},
  \bibinfo {author} {\bibfnamefont {E.~D.}\ \bibnamefont {Kuhnle}}, \bibinfo
  {author} {\bibfnamefont {Y.}~\bibnamefont {Wang}}, \bibinfo {author}
  {\bibfnamefont {C.~H.}\ \bibnamefont {Greene}},\ and\ \bibinfo {author}
  {\bibfnamefont {M.}~\bibnamefont {Weidem\"uller}},\ }\href
  {https://journals.aps.org/pra/abstract/10.1103/PhysRevA.95.062708} {\bibfield
   {journal} {\bibinfo  {journal} {Phys. Rev. A}\ }\textbf {\bibinfo {volume}
  {95}},\ \bibinfo {pages} {062708} (\bibinfo {year} {2017})}\BibitemShut
  {NoStop}%
\bibitem [{\citenamefont {Johansen}\ \emph {et~al.}(2017)\citenamefont
  {Johansen}, \citenamefont {DeSalvo}, \citenamefont {Patel},\ and\
  \citenamefont {Chin}}]{Johansen2017tuo}%
  \BibitemOpen
  \bibfield  {author} {\bibinfo {author} {\bibfnamefont {J.}~\bibnamefont
  {Johansen}}, \bibinfo {author} {\bibfnamefont {B.}~\bibnamefont {DeSalvo}},
  \bibinfo {author} {\bibfnamefont {K.}~\bibnamefont {Patel}},\ and\ \bibinfo
  {author} {\bibfnamefont {C.}~\bibnamefont {Chin}},\ }\href
  {https://www.nature.com/articles/nphys4130} {\bibfield  {journal} {\bibinfo
  {journal} {Nat. Phys.}\ }\textbf {\bibinfo {volume} {13}},\ \bibinfo {pages}
  {731} (\bibinfo {year} {2017})}\BibitemShut {NoStop}%
\bibitem [{\citenamefont {Jag}\ \emph {et~al.}(2014)\citenamefont {Jag},
  \citenamefont {Zaccanti}, \citenamefont {Cetina}, \citenamefont {Lous},
  \citenamefont {Schreck}, \citenamefont {Grimm}, \citenamefont {Petrov},\ and\
  \citenamefont {Levinsen}}]{Jag2014ooa}%
  \BibitemOpen
  \bibfield  {author} {\bibinfo {author} {\bibfnamefont {M.}~\bibnamefont
  {Jag}}, \bibinfo {author} {\bibfnamefont {M.}~\bibnamefont {Zaccanti}},
  \bibinfo {author} {\bibfnamefont {M.}~\bibnamefont {Cetina}}, \bibinfo
  {author} {\bibfnamefont {R.~S.}\ \bibnamefont {Lous}}, \bibinfo {author}
  {\bibfnamefont {F.}~\bibnamefont {Schreck}}, \bibinfo {author} {\bibfnamefont
  {R.}~\bibnamefont {Grimm}}, \bibinfo {author} {\bibfnamefont {D.~S.}\
  \bibnamefont {Petrov}},\ and\ \bibinfo {author} {\bibfnamefont
  {J.}~\bibnamefont {Levinsen}},\ }\href
  {https://doi.org/10.1103/PhysRevLett.112.075302} {\bibfield  {journal}
  {\bibinfo  {journal} {Phys. Rev. Lett.}\ }\textbf {\bibinfo {volume} {112}},\
  \bibinfo {pages} {075302} (\bibinfo {year} {2014})}\BibitemShut {NoStop}%
\bibitem [{\citenamefont {Huang}(2020)}]{Huang2020bec}%
  \BibitemOpen
  \bibfield  {author} {\bibinfo {author} {\bibfnamefont {B.}~\bibnamefont
  {Huang}},\ }\href
  {https://journals.aps.org/pra/abstract/10.1103/PhysRevA.101.063618}
  {\bibfield  {journal} {\bibinfo  {journal} {Phys. Rev. A}\ }\textbf {\bibinfo
  {volume} {101}},\ \bibinfo {pages} {063618} (\bibinfo {year}
  {2020})}\BibitemShut {NoStop}%
\bibitem [{\citenamefont {Ospelkaus}\ \emph {et~al.}(2006)\citenamefont
  {Ospelkaus}, \citenamefont {Ospelkaus}, \citenamefont {Sengstock},\ and\
  \citenamefont {Bongs}}]{Ospelkaus2006idd}%
  \BibitemOpen
  \bibfield  {author} {\bibinfo {author} {\bibfnamefont {C.}~\bibnamefont
  {Ospelkaus}}, \bibinfo {author} {\bibfnamefont {S.}~\bibnamefont
  {Ospelkaus}}, \bibinfo {author} {\bibfnamefont {K.}~\bibnamefont
  {Sengstock}},\ and\ \bibinfo {author} {\bibfnamefont {K.}~\bibnamefont
  {Bongs}},\ }\href {https://doi.org/10.1103/PhysRevLett.96.020401} {\bibfield
  {journal} {\bibinfo  {journal} {Phys. Rev. Lett.}\ }\textbf {\bibinfo
  {volume} {96}},\ \bibinfo {eid} {020401} (\bibinfo {year}
  {2006})}\BibitemShut {NoStop}%
\bibitem [{\citenamefont {Zaccanti}\ \emph {et~al.}(2006)\citenamefont
  {Zaccanti}, \citenamefont {D'Errico}, \citenamefont {Ferlaino}, \citenamefont
  {Roati}, \citenamefont {Inguscio},\ and\ \citenamefont
  {Modugno}}]{Zaccanti2006cot}%
  \BibitemOpen
  \bibfield  {author} {\bibinfo {author} {\bibfnamefont {M.}~\bibnamefont
  {Zaccanti}}, \bibinfo {author} {\bibfnamefont {C.}~\bibnamefont {D'Errico}},
  \bibinfo {author} {\bibfnamefont {F.}~\bibnamefont {Ferlaino}}, \bibinfo
  {author} {\bibfnamefont {G.}~\bibnamefont {Roati}}, \bibinfo {author}
  {\bibfnamefont {M.}~\bibnamefont {Inguscio}},\ and\ \bibinfo {author}
  {\bibfnamefont {G.}~\bibnamefont {Modugno}},\ }\href
  {https://doi.org/10.1103/PhysRevA.74.041605} {\bibfield  {journal} {\bibinfo
  {journal} {Physical Review A}\ }\textbf {\bibinfo {volume} {74}},\ \bibinfo
  {pages} {041605} (\bibinfo {year} {2006})}\BibitemShut {NoStop}%
\bibitem [{\citenamefont {Spiegelhalder}\ \emph {et~al.}(2010)\citenamefont
  {Spiegelhalder}, \citenamefont {Trenkwalder}, \citenamefont {Naik},
  \citenamefont {Kerner}, \citenamefont {Wille}, \citenamefont {Hendl},
  \citenamefont {Schreck},\ and\ \citenamefont {Grimm}}]{Spiegelhalder2010aop}%
  \BibitemOpen
  \bibfield  {author} {\bibinfo {author} {\bibfnamefont {F.~M.}\ \bibnamefont
  {Spiegelhalder}}, \bibinfo {author} {\bibfnamefont {A.}~\bibnamefont
  {Trenkwalder}}, \bibinfo {author} {\bibfnamefont {D.}~\bibnamefont {Naik}},
  \bibinfo {author} {\bibfnamefont {G.}~\bibnamefont {Kerner}}, \bibinfo
  {author} {\bibfnamefont {E.}~\bibnamefont {Wille}}, \bibinfo {author}
  {\bibfnamefont {G.}~\bibnamefont {Hendl}}, \bibinfo {author} {\bibfnamefont
  {F.}~\bibnamefont {Schreck}},\ and\ \bibinfo {author} {\bibfnamefont
  {R.}~\bibnamefont {Grimm}},\ }\href
  {https://doi.org/10.1103/PhysRevA.81.043637} {\bibfield  {journal} {\bibinfo
  {journal} {Phys. Rev. A}\ }\textbf {\bibinfo {volume} {81}},\ \bibinfo
  {pages} {043637} (\bibinfo {year} {2010})}\BibitemShut {NoStop}%
\bibitem [{\citenamefont {Lous}(2018)}]{Lous2018PHD}%
  \BibitemOpen
  \bibfield  {author} {\bibinfo {author} {\bibfnamefont {R.~S.}\ \bibnamefont
  {Lous}},\ }\emph {\bibinfo {title} {Tunable {{Bose}}-{{Fermi}} and
  {{Fermi}}-{{Fermi Mixtures}} of {{Potassium}} and {{Lithium}}: {{Phase
  Separation}}, {{Polarons}}, and {{Molecules}}}},\ \href@noop {} {Ph.D.
  thesis},\ \bibinfo  {school} {Innsbruck University} (\bibinfo {year}
  {2018})\BibitemShut {NoStop}%
\bibitem [{Note1()}]{Note1}%
  \BibitemOpen
  \bibinfo {note} {Note that the atom numbers in the PBEC and in the THC
  slightly differ, due to the different preparation methods.}\BibitemShut
  {Stop}%
\bibitem [{Han()}]{HanTie}%
  \BibitemOpen
  \href@noop {} {}\bibinfo {note} {Hanna, T. and Tiesinga, E.(private
  communication)}\BibitemShut {NoStop}%
\bibitem [{\citenamefont {Petrov}(2004)}]{Petrov2004tbp}%
  \BibitemOpen
  \bibfield  {author} {\bibinfo {author} {\bibfnamefont {D.~S.}\ \bibnamefont
  {Petrov}},\ }\href {https://doi.org/10.1103/PhysRevLett.93.143201} {\bibfield
   {journal} {\bibinfo  {journal} {Phys. Rev. Lett.}\ }\textbf {\bibinfo
  {volume} {93}},\ \bibinfo {pages} {143201} (\bibinfo {year}
  {2004})}\BibitemShut {NoStop}%
\bibitem [{Not()}]{NoteIF22}%
  \BibitemOpen
  \href@noop {} {}\bibinfo {note} {The mean density of the condensed part of
  the partial BEC amounts to 4/7 of the peak density, in the Thomas Fermi
  approximation}\BibitemShut {NoStop}%
\bibitem [{\citenamefont {Liu}\ \emph {et~al.}(2020)\citenamefont {Liu},
  \citenamefont {Shi}, \citenamefont {Parish},\ and\ \citenamefont
  {Levinsen}}]{Liu2020}%
  \BibitemOpen
  \bibfield  {author} {\bibinfo {author} {\bibfnamefont {W.~E.}\ \bibnamefont
  {Liu}}, \bibinfo {author} {\bibfnamefont {Z.}~\bibnamefont {Shi}}, \bibinfo
  {author} {\bibfnamefont {M.~M.}\ \bibnamefont {Parish}},\ and\ \bibinfo
  {author} {\bibfnamefont {J.}~\bibnamefont {Levinsen}},\ }\href
  {https://doi.org/https://doi.org/10.1103/PhysRevA.102.023304} {\bibfield
  {journal} {\bibinfo  {journal} {Physical Review A}\ }\textbf {\bibinfo
  {volume} {102}},\ \bibinfo {pages} {023304} (\bibinfo {year}
  {2020})}\BibitemShut {NoStop}%
\bibitem [{\citenamefont {Massignan}\ and\ \citenamefont
  {Bruun}(2011)}]{Massignan2011}%
  \BibitemOpen
  \bibfield  {author} {\bibinfo {author} {\bibfnamefont {P.}~\bibnamefont
  {Massignan}}\ and\ \bibinfo {author} {\bibfnamefont {G.~M.}\ \bibnamefont
  {Bruun}},\ }\href {https://doi.org/10.1140/epjd/e2011-20084-5} {\bibfield
  {journal} {\bibinfo  {journal} {Eur. Phys. J. D}\ }\textbf {\bibinfo {volume}
  {65}},\ \bibinfo {pages} {83} (\bibinfo {year} {2011})}\BibitemShut {NoStop}%
\bibitem [{not()}]{noteIF20}%
  \BibitemOpen
  \href@noop {} {}\bibinfo {note} {Attributed to imaging artefacts, our
  absorption pictures show a residual signal in K$\ket{1}$ when imaging a BEC
  in K$\ket{2}$. We explain this by scattering off-resonant light from a very
  dense atomic sample and subtract the resulting signal from the real atom
  number.}\BibitemShut {Stop}%
\bibitem [{Tie()}]{TiemannPrievComm}%
  \BibitemOpen
  \href@noop {} {}\bibinfo {note} {Eberhard Tiemann. (private
  communication)}\BibitemShut {NoStop}%
\bibitem [{\citenamefont {Petrov}(2003)}]{Petrov2003tbp}%
  \BibitemOpen
  \bibfield  {author} {\bibinfo {author} {\bibfnamefont {D.~S.}\ \bibnamefont
  {Petrov}},\ }\href {https://doi.org/10.1103/PhysRevA.67.010703} {\bibfield
  {journal} {\bibinfo  {journal} {Physical Review A}\ }\textbf {\bibinfo
  {volume} {67}},\ \bibinfo {pages} {010703} (\bibinfo {year}
  {2003})}\BibitemShut {NoStop}%
\bibitem [{\citenamefont {Levinsen}\ and\ \citenamefont
  {Petrov}(2011)}]{Levinsen2011ada}%
  \BibitemOpen
  \bibfield  {author} {\bibinfo {author} {\bibfnamefont {J.}~\bibnamefont
  {Levinsen}}\ and\ \bibinfo {author} {\bibfnamefont {D.}~\bibnamefont
  {Petrov}},\ }\href {https://doi.org/10.1140/epjd/e2011-20071-x} {\bibfield
  {journal} {\bibinfo  {journal} {Eur. Phys. J. D}\ }\textbf {\bibinfo {volume}
  {65}},\ \bibinfo {pages} {67} (\bibinfo {year} {2011})}\BibitemShut {NoStop}%
\bibitem [{\citenamefont {Chin}\ and\ \citenamefont
  {Julienne}(2005)}]{Chin2005rft}%
  \BibitemOpen
  \bibfield  {author} {\bibinfo {author} {\bibfnamefont {C.}~\bibnamefont
  {Chin}}\ and\ \bibinfo {author} {\bibfnamefont {P.~S.}\ \bibnamefont
  {Julienne}},\ }\href
  {https://doi.org/https://doi.org/10.1103/PhysRevA.71.012713} {\bibfield
  {journal} {\bibinfo  {journal} {Phys. Rev. A}\ }\textbf {\bibinfo {volume}
  {71}},\ \bibinfo {pages} {012713} (\bibinfo {year} {2005})}\BibitemShut
  {NoStop}%
\bibitem [{\citenamefont {Combescot}\ \emph {et~al.}(2007)\citenamefont
  {Combescot}, \citenamefont {Recati}, \citenamefont {Lobo},\ and\
  \citenamefont {Chevy}}]{Combescot2007nso}%
  \BibitemOpen
  \bibfield  {author} {\bibinfo {author} {\bibfnamefont {R.}~\bibnamefont
  {Combescot}}, \bibinfo {author} {\bibfnamefont {A.}~\bibnamefont {Recati}},
  \bibinfo {author} {\bibfnamefont {C.}~\bibnamefont {Lobo}},\ and\ \bibinfo
  {author} {\bibfnamefont {F.}~\bibnamefont {Chevy}},\ }\href
  {https://doi.org/10.1103/PhysRevLett.98.180402} {\bibfield  {journal}
  {\bibinfo  {journal} {Phys. Rev. Lett.}\ }\textbf {\bibinfo {volume} {98}},\
  \bibinfo {pages} {180402} (\bibinfo {year} {2007})}\BibitemShut {NoStop}%
\bibitem [{\citenamefont {Prokof'ev}\ and\ \citenamefont
  {Svistunov}(2008)}]{Prokofev2008}%
  \BibitemOpen
  \bibfield  {author} {\bibinfo {author} {\bibfnamefont {N.}~\bibnamefont
  {Prokof'ev}}\ and\ \bibinfo {author} {\bibfnamefont {B.}~\bibnamefont
  {Svistunov}},\ }\href {https://doi.org/10.1103/PhysRevB.77.020408} {\bibfield
   {journal} {\bibinfo  {journal} {Physical Review B}\ }\textbf {\bibinfo
  {volume} {77}},\ \bibinfo {pages} {020408} (\bibinfo {year}
  {2008})}\BibitemShut {NoStop}%
\bibitem [{\citenamefont {Koschorreck}\ \emph {et~al.}(2012)\citenamefont
  {Koschorreck}, \citenamefont {Pertot}, \citenamefont {Vogt}, \citenamefont
  {Fr{\ifmmode\ddot{o}\else\"{o}\fi}hlich}, \citenamefont {Feld},\ and\
  \citenamefont {K{\ifmmode\ddot{o}\else\"{o}\fi}hl}}]{Koschorreck2012}%
  \BibitemOpen
  \bibfield  {author} {\bibinfo {author} {\bibfnamefont {M.}~\bibnamefont
  {Koschorreck}}, \bibinfo {author} {\bibfnamefont {D.}~\bibnamefont {Pertot}},
  \bibinfo {author} {\bibfnamefont {E.}~\bibnamefont {Vogt}}, \bibinfo {author}
  {\bibfnamefont {B.}~\bibnamefont {Fr{\ifmmode\ddot{o}\else\"{o}\fi}hlich}},
  \bibinfo {author} {\bibfnamefont {M.}~\bibnamefont {Feld}},\ and\ \bibinfo
  {author} {\bibfnamefont {M.}~\bibnamefont
  {K{\ifmmode\ddot{o}\else\"{o}\fi}hl}},\ }\href
  {https://doi.org/10.1038/nature11151} {\bibfield  {journal} {\bibinfo
  {journal} {Nature}\ }\textbf {\bibinfo {volume} {485}},\ \bibinfo {pages}
  {619} (\bibinfo {year} {2012})}\BibitemShut {NoStop}%
\bibitem [{\citenamefont {Mora}\ and\ \citenamefont
  {Chevy}(2009)}]{Mora2009gso}%
  \BibitemOpen
  \bibfield  {author} {\bibinfo {author} {\bibfnamefont {C.}~\bibnamefont
  {Mora}}\ and\ \bibinfo {author} {\bibfnamefont {F.}~\bibnamefont {Chevy}},\
  }\href {https://doi.org/https://doi.org/10.1103/PhysRevA.80.033607}
  {\bibfield  {journal} {\bibinfo  {journal} {Phys. Rev. A}\ }\textbf {\bibinfo
  {volume} {80}},\ \bibinfo {pages} {033607} (\bibinfo {year}
  {2009})}\BibitemShut {NoStop}%
\bibitem [{\citenamefont {Punk}\ \emph {et~al.}(2009)\citenamefont {Punk},
  \citenamefont {Dumitrescu},\ and\ \citenamefont {Zwerger}}]{Punk2009}%
  \BibitemOpen
  \bibfield  {author} {\bibinfo {author} {\bibfnamefont {M.}~\bibnamefont
  {Punk}}, \bibinfo {author} {\bibfnamefont {P.~T.}\ \bibnamefont
  {Dumitrescu}},\ and\ \bibinfo {author} {\bibfnamefont {W.}~\bibnamefont
  {Zwerger}},\ }\href {https://doi.org/10.1103/PhysRevA.80.053605} {\bibfield
  {journal} {\bibinfo  {journal} {Phys. Rev. A}\ }\textbf {\bibinfo {volume}
  {80}},\ \bibinfo {pages} {053605} (\bibinfo {year} {2009})}\BibitemShut
  {NoStop}%
\bibitem [{\citenamefont {Combescot}\ \emph {et~al.}(2009)\citenamefont
  {Combescot}, \citenamefont {Giraud},\ and\ \citenamefont
  {Leyronas}}]{Combescot2009}%
  \BibitemOpen
  \bibfield  {author} {\bibinfo {author} {\bibfnamefont {R.}~\bibnamefont
  {Combescot}}, \bibinfo {author} {\bibfnamefont {S.}~\bibnamefont {Giraud}},\
  and\ \bibinfo {author} {\bibfnamefont {X.}~\bibnamefont {Leyronas}},\ }\href
  {https://doi.org/10.1209/0295-5075/88/60007} {\bibfield  {journal} {\bibinfo
  {journal} {{EPL} (Europhysics Letters)}\ }\textbf {\bibinfo {volume} {88}},\
  \bibinfo {pages} {60007} (\bibinfo {year} {2009})}\BibitemShut {NoStop}%
\bibitem [{\citenamefont {Trefzger}\ and\ \citenamefont
  {Castin}(2012)}]{Trefzger2012}%
  \BibitemOpen
  \bibfield  {author} {\bibinfo {author} {\bibfnamefont {C.}~\bibnamefont
  {Trefzger}}\ and\ \bibinfo {author} {\bibfnamefont {Y.}~\bibnamefont
  {Castin}},\ }\href {https://doi.org/10.1103/PhysRevA.85.053612} {\bibfield
  {journal} {\bibinfo  {journal} {Phys. Rev. A}\ }\textbf {\bibinfo {volume}
  {85}},\ \bibinfo {pages} {053612} (\bibinfo {year} {2012})}\BibitemShut
  {NoStop}%
\bibitem [{\citenamefont {Qi}\ and\ \citenamefont {Zhai}(2012)}]{Qi2012hpf}%
  \BibitemOpen
  \bibfield  {author} {\bibinfo {author} {\bibfnamefont {R.}~\bibnamefont
  {Qi}}\ and\ \bibinfo {author} {\bibfnamefont {H.}~\bibnamefont {Zhai}},\
  }\href {https://doi.org/10.1103/PhysRevA.85.041603} {\bibfield  {journal}
  {\bibinfo  {journal} {Phys. Rev. A}\ }\textbf {\bibinfo {volume} {85}},\
  \bibinfo {pages} {041603(R)} (\bibinfo {year} {2012})}\BibitemShut {NoStop}%
\bibitem [{\citenamefont {Bruun}\ and\ \citenamefont
  {Massignan}(2010)}]{Bruun2010}%
  \BibitemOpen
  \bibfield  {author} {\bibinfo {author} {\bibfnamefont {G.~M.}\ \bibnamefont
  {Bruun}}\ and\ \bibinfo {author} {\bibfnamefont {P.}~\bibnamefont
  {Massignan}},\ }\href {https://doi.org/10.1103/PhysRevLett.105.020403}
  {\bibfield  {journal} {\bibinfo  {journal} {Phys. Rev. Lett.}\ }\textbf
  {\bibinfo {volume} {105}},\ \bibinfo {pages} {020403} (\bibinfo {year}
  {2010})}\BibitemShut {NoStop}%
\end{thebibliography}%
\end{document}